\pdfoutput=1
\documentclass[a4paper,11pt]{article}
\usepackage{jheppub}
\makeatletter
\renewcommand{\@fpheader}{}
\makeatother

\usepackage[T1]{fontenc}
\usepackage[utf8]{inputenc}
\usepackage{graphicx}
\usepackage{hyperref}
\hypersetup{colorlinks=true,linkcolor=blue,citecolor=blue,urlcolor=blue}
\usepackage{array}
\usepackage{booktabs}
\usepackage{makecell}
\usepackage{multirow}

\newcommand{\widthEr}{\linewidth}
\newcommand{\widthVrin}{\linewidth}
\newcommand{\widthVrout}{\linewidth}
\newcommand{\widthLin}{\linewidth}
\newcommand{\widthLout}{\linewidth}
\newcommand{\widthLsin}{\linewidth}
\newcommand{\widthLsout}{\linewidth}
\newcommand{\widthLsoutOne}{\linewidth}
\newcommand{\widthSa}{\linewidth}
\newcommand{\widthSainout}{\linewidth}
\newcommand{\widthSain}{\linewidth}
\newcommand{\widthSaout}{\linewidth}
\newcommand{\widthSaoutOne}{\linewidth}
\newcommand{\widthLainZero}{\linewidth}
\newcommand{\widthLainPos}{\linewidth}
\newcommand{\widthLainNeg}{\linewidth}
\newcommand{\widthLaoutZero}{\linewidth}
\newcommand{\widthLaoutPos}{\linewidth}
\newcommand{\widthLaoutNeg}{\linewidth}

\newcommand{\heightEr}{6cm}
\newcommand{\heightVrin}{5cm}
\newcommand{\heightVrout}{5cm}
\newcommand{\heightLin}{6.2cm}
\newcommand{\heightLout}{6.2cm}
\newcommand{\heightLsin}{5cm}
\newcommand{\heightLsout}{5cm}
\newcommand{\heightLsoutOne}{5cm}
\newcommand{\heightSa}{6.2cm}
\newcommand{\heightSainout}{6.2cm}
\newcommand{\heightSain}{5.2cm}
\newcommand{\heightSaout}{5.2cm}
\newcommand{\heightSaoutOne}{5.2cm}
\newcommand{\heightLainZero}{5.2cm}
\newcommand{\heightLainPos}{5.2cm}
\newcommand{\heightLainNeg}{5.2cm}
\newcommand{\heightLaoutZero}{5.2cm}
\newcommand{\heightLaoutPos}{5.2cm}
\newcommand{\heightLaoutNeg}{5.2cm}
\newcommand{\panelcaption}[1]{\par\smallskip{\small #1}}

\title{Chaos bound for spinning particles in a hairy black hole with two photon spheres}

\author{Chuanhong Gao}
\emailAdd{chuanhonggao@hotmail.com}

\affiliation{Sichuan Technology and Business University,
Chengdu 611745, China}

\abstract{
In this work, we investigate the unstable circular orbits of spinning test particles and the associated classical orbital chaos bound in a hairy black hole spacetime admitting a double photon sphere structure. This background supports two distinct branches of unstable circular orbits, referred to as the inner and outer branches. We find that, within the parameter ranges considered, the inner branch always satisfies the chaos bound, whereas violations can occur on the outer branch. This behavior contrasts with many previous studies focusing primarily on near-horizon violations of the chaos bound and reveals a clear dependence of the bound on the orbital branch. Our analysis further shows that the total angular momentum of the particle has a pronounced effect on the local orbital instability and can enlarge the bound-violating region on the outer branch. Meanwhile, the deformation parameter of the background geometry regulates the validity of the bound by modifying both the orbital Lyapunov exponent and the black hole surface gravity. By contrast, the effect of the particle spin is comparatively weak and mainly shifts the existing unstable circular orbits and the corresponding critical boundaries through the spin-curvature coupling.
}


\begin{document}
\maketitle
\flushbottom

\section{Introduction}
\label{sec:introduction}

The relationship between the growth rate of chaos and black hole temperature is an important topic in studies of quantum chaos and black hole dynamics~\cite{Shenker:2013pqa,Shenker:2013yza,Shenker:2014cwa}. The chaos bound proposed by Maldacena, Shenker and Stanford (MSS) states that, for thermal quantum systems with a large number of degrees of freedom, the Lyapunov exponent characterized by out-of-time-order correlators (OTOCs) is bounded in terms of the temperature~\cite{Maldacena:2015waa}, namely
\begin{equation}
  \lambda\leq \frac{2\pi k_B T}{\hbar}.
  \label{eq:mss-chaos-bound}
\end{equation}
where $\lambda$ is the Lyapunov exponent, $T$ is the temperature, $k_B$ is the Boltzmann constant, and $\hbar$ is the reduced Planck constant. Because the black hole temperature is proportional to the horizon surface gravity, it is natural to ask whether the horizon scale also constrains the instability of classical particle orbits. Hashimoto and Tanahashi studied particle motion near the horizons of static and spherically symmetric black holes and showed that the Lyapunov exponent associated with a near-horizon unstable equilibrium is universally determined by the surface gravity~\cite{Hashimoto:2016dfz}. In particular, they showed that the near-horizon Lyapunov exponent is governed by the black hole surface gravity $\kappa$ and, in the absence of additional sources of chaos, obtained the bound
\begin{equation}
\lambda\leq\kappa .
\label{eq:surface-gravity-chaos-bound}
\end{equation}
This relation motivates the comparison between the orbital Lyapunov exponent and the black hole surface gravity as a useful diagnostic for examining the interplay between classical orbital instability and the horizon scale~\cite{Hashimoto:2022bll,Hashimoto:2023mgl,Gallo:2025light}.

Subsequent studies have examined this relation in a variety of black hole backgrounds and test particle systems. Zhao et al. showed that, for charged particles in static equilibrium near charged black holes, the Lyapunov exponent approaches the surface gravity as the equilibrium point approaches the horizon, whereas its behavior at finite radius depends on the background geometry~\cite{Zhao:2018wkl}. Lei and Ge extended the analysis to circular motion of charged particles and showed that the orbital angular momentum shifts the circular orbit radius and modifies the curvature of the effective potential, thereby affecting the relation between the Lyapunov exponent and the surface gravity~\cite{Lei:2021koj,Lei:2022hzg}. The analysis was subsequently extended to Reissner–Nordström (RN), RN–anti-de Sitter (RN–AdS), Kerr–Newman, Kerr–Newman–de Sitter (dS), and Kerr–Newman–AdS black holes~\cite{Lei:2024rn,Kan:2021blg,Gwak:2022xje,Park:2024jhg}. Black brane backgrounds and black hole backgrounds with higher-curvature corrections have also been considered~\cite{Lei:2023hyperscaling,Dutta:2025pbrane,Xie:2023gbads,Das:2024nearhorizon}. Further investigations of chaos bound behavior in black hole spacetimes are reported in Refs.~\cite{Gao:2022ybx,Chen:2022dpi,Yu:2023taubnut,Yu:2022emda,Lee:2025kerrsen}. Taken together, these results indicate that, away from the near-horizon regime, the relation between the orbital Lyapunov exponent and the surface gravity is sensitive to both particle properties and the background geometry. Electromagnetic interactions, orbital angular momentum, black hole rotation, the cosmological constant, and higher-curvature corrections can modify the effective radial dynamics and thereby alter the relative magnitude of $\lambda$ and $\kappa$, allowing $\lambda>\kappa$ in appropriate regions of parameter space.

Spinning particles provide an additional probe of this relation. Their dynamics is governed by the Mathisson–Papapetrou–Dixon (MPD) equations, in which the intrinsic spin couples to the background curvature. This spin-curvature interaction modifies the radial dynamics, shifting the circular orbit locations and altering their local stability. Consequently, the corresponding Lyapunov exponent depends not only on the orbital parameters but also on the magnitude and orientation of the particle spin. Previous studies have shown that particle spin, total angular momentum, and background geometry can jointly modify the relation between the orbital Lyapunov exponent and the surface gravity~\cite{Yang:2026rnspin,Li:2026gbspin,Li:2026eehspin,Yang:2026lvspin}. Spin-curvature coupling therefore provides an additional mechanism for modifying orbital instability and the associated chaos bound behavior.

Black holes admitting two photon spheres provide a natural setting for investigating multibranch orbital instability. In the configurations of interest, the photon effective potential develops two local maxima outside the event horizon, corresponding to two unstable photon spheres and indicating a richer radial structure than in the single photon sphere case~\cite{Tsukamoto:2021caq,Guo:2022umh}. Such multiple photon sphere configurations have been shown to produce distinctive features in photon motion and optical observables, including black hole images and strong gravitational lensing~\cite{Gan:2021xdl,Gan:2021pwu,Wang:2025doublephoton}. This richer radial structure raises the possibility that massive particle circular orbits located on different radial branches may exhibit distinct local instabilities and hence different Lyapunov exponents. For spinning particles, spin-curvature coupling further shifts the circular orbit locations and modifies the local instability on each branch, while the total angular momentum provides an additional control parameter for the orbital dynamics. A concrete realization is the hairy Schwarzschild black hole (hSBH), constructed from the Schwarzschild seed geometry through the gravitational decoupling approach with an additional matter source~\cite{Ovalle:2020kpd}. In a finite region of parameter space, this geometry can support two unstable photon spheres outside the event horizon~\cite{Guo:2022umh}. Previous investigations of hairy black holes and multiple photon sphere configurations have mainly addressed photon motion, black hole shadows, strong gravitational lensing, interferometric signatures, and quasinormal modes~\cite{Guo:2021bwr,Guo:2022blz,Chen:2023interferometric}. By contrast, the instability and chaos bound behavior of massive spinning particles on the coexisting circular orbit branches of this background remain largely unexplored.

In this paper, we study the radial instability of spinning particles in the hSBH admitting two photon spheres and examine the associated classical orbital chaos bound. The MPD equations supplemented by the Tulczyjew--Dixon condition are used to determine the unstable circular orbit branches, and the corresponding orbital Lyapunov exponents are compared with the black hole surface gravity. Our analysis addresses three related questions: whether the double photon sphere geometry supports multiple unstable circular orbit branches for spinning particles, whether the inner and outer branches exhibit distinct Lyapunov exponent and chaos bound behaviors, and how the particle spin, total angular momentum, and background deformation parameter influence the orbital instability and the critical boundary separating the bound-satisfying and bound-violating regimes.

The paper is organized as follows. Section~\ref{sec:double-photon-sphere-dynamics} introduces the hSBH background, the MPD dynamics of spinning particles, and the Lyapunov exponent prescription. Section~\ref{sec:chaos-bound-violation} presents the numerical results for the inner and outer branches and analyzes the dependence on the particle and background parameters. Section~\ref{sec:conclusions} summarizes the main results and discusses their physical implications.

\section{Dynamics of double photon sphere black holes}
\label{sec:double-photon-sphere-dynamics}

\subsection{The double photon sphere black hole solution}
\label{subsec:double-photon-sphere-solution}

We consider the hSBH obtained through the gravitational decoupling approach~\cite{Ovalle:2020kpd}. Since the construction of this solution has been discussed in detail in Ref.~\cite{Ovalle:2020kpd}, here we take the resulting geometry as the background spacetime without repeating its derivation. The resulting hSBH is described by
\begin{equation}
  ds^2=-f(r)dt^2+\frac{dr^2}{f(r)}
  +r^2\left(d\theta^2+\sin^2\theta\,d\phi^2\right),
  \label{eq:hairy-schwarzschild-metric}
\end{equation}
where
\begin{equation}
  f(r)=1-\frac{2M}{r}
  +\alpha e^{-\frac{r}{M-l_0/2}} .
  \label{eq:hairy-schwarzschild-f}
\end{equation}
In this expression, $M$ is the black hole mass, $\alpha$ is the dimensionless deformation parameter, and $l_0$ is the primary hair charge. The primary hair charge is related to a length parameter $l$ by
 $l_0=\alpha l$. The condition $l_0\leq 2M$ is imposed in the original construction to preserve the asymptotic flatness of the spacetime. The event horizon radius $r_h$ is determined by $f(r_h)=0$. For $\alpha=0$, the metric function reduces to $f(r)=1-2M/r$, and the standard Schwarzschild solution is recovered.

\noindent For the metric in eq.~\eqref{eq:hairy-schwarzschild-metric}, the surface gravity at the event horizon is obtained from $\kappa = \frac{1}{2} f'(r_h)$, which gives
\begin{equation}
  \kappa
  =\frac{M}{r_h^2}
  -\frac{\alpha}{2(M-l_0/2)}
  e^{-\frac{r_h}{M-l_0/2}} .
  \label{eq:surface-gravity-hairy-schwarzschild}
\end{equation}
This surface gravity will be used as the reference scale in the chaos bound relation.

\subsection{Equations of motion for spinning particles}
\label{subsec:spinning-particle-equations}

The motion of a spinning particle in a curved background is described by the MPD equations~\cite{Hojman1977},
\begin{equation}
\begin{gathered}
  \frac{D\tilde p^{\mu}}{D\tau}
  =-\frac{1}{2}R^\mu{}_{\nu\alpha\beta}
  u^\nu \tilde S^{\alpha\beta},
  \\[3pt]
  \frac{D\tilde S^{\mu\nu}}{D\tau}
  =\tilde p^{\mu}u^{\nu}
  -u^{\mu}\tilde p^{\nu}.
  \end{gathered}
  \label{eq:mpd-equations}
\end{equation}
Here, $\tilde p^\mu$ is the four momentum, $u^\mu=dx^\mu/d\tau$ is the four velocity, $\tilde S^{\mu\nu}$ is the antisymmetric spin tensor, and $R^\mu{}_{\nu\alpha\beta}$ denotes the Riemann tensor of the background spacetime. The covariant derivative is taken along the particle worldline. This formalism has been widely used to describe the dynamics of spinning test bodies in curved spacetime, including cases with electromagnetic interactions and spin-gravity coupling~\cite{Hojman1977,Wald:1972sz}. Since the MPD equations do not by themselves determine a unique representative worldline for an extended spinning body, a spin supplementary condition must be imposed. In this work, we adopt the Tulczyjew--Dixon(TD) supplementary condition~\cite{Tulczyjew:1959,Hanson:1974qy},
\begin{equation}
  \tilde S^{\mu\nu}\tilde p_{\nu}=0 .
  \label{eq:tulczyjew-dixon-condition}
\end{equation}

It is convenient to work with quantities per unit mass. We introduce
\begin{equation}
  p^\mu=\frac{\tilde p^\mu}{m},
  \qquad
  S^{\mu\nu}=\frac{\tilde S^{\mu\nu}}{m},
  \qquad
  s=\pm\frac{\tilde S}{m},
  \label{eq:unit-mass-variables}
\end{equation}
where $m$ is the particle mass and $\tilde S$ is the magnitude of the spin. The sign of $s$ specifies the orientation of the particle spin relative to its angular momentum: $s>0$ corresponds to the aligned configuration, whereas $s<0$ corresponds to the anti-aligned configuration. The normalized momentum and spin satisfy
\begin{equation}
  p^\mu p_\mu=-1, 
  \label{eq:momentum-normalization-1}
\end{equation}
\begin{equation}
  s^2=\frac{1}{2}S_{\mu\nu}S^{\mu\nu}.
  \label{eq:momentum-normalization-2}
\end{equation}

For the background metric \eqref{eq:hairy-schwarzschild-metric} considered in this work, we restrict the motion to the equatorial plane, $\theta=\frac{\pi}{2}$, and choose the spin direction to be orthogonal to the orbital plane. Under this configuration, $p^\theta=0$, $u^\theta=0$ and $S^{\theta\mu}=0$. The nonzero independent components of the spin tensor are then $S^{tr}$, $S^{t\phi}$ and $S^{r\phi}$. Under the TD condition, these spin components are not independent. For the metric considered here, the condition $S^{\mu\nu}p_\nu=0$ gives
\begin{equation}
  \frac{p^r}{f(r)}S^{tr}
  +r^2p^\phi S^{t\phi}=0,
  \label{eq:td-component-t}
\end{equation}
\begin{equation}
  r^2p^\phi S^{r\phi}
  +f(r)p^t S^{tr}=0,
  \label{eq:td-component-r}
\end{equation}
\begin{equation}
  \frac{p^r}{f(r)}S^{r\phi}
  -f(r)p^t S^{t\phi}=0.
  \label{eq:td-component-phi}
\end{equation}
Together with the spin invariant, these relations allow the spin tensor to be written in terms of the particle momentum and the spin parameter $s$ as in the treatment of spinning massive test particles in general static spherically symmetric spacetimes~\cite{Zalaquett:2014koa},
\begin{equation}
  S^{tr}=-\frac{s p_\phi}{r},
  \qquad
  S^{t\phi}=\frac{s p_r}{r},
  \qquad
  S^{r\phi}=-\frac{s p_t}{r}.
  \label{eq:spin-components-momentum}
\end{equation}
These relations express the independent spin tensor components in terms of the momentum variables and the spin parameter.

To investigate the orbital dynamics of a spinning particle around the black hole, we follow the approach developed in ~\cite{Yang:2026rnspin,Zalaquett:2014koa}. The stationarity and axial symmetry of the spacetime are associated with the Killing vectors $(\partial/\partial t)^a$ and $(\partial/\partial \phi)^a$, respectively. For a spinning particle, the conserved quantities associated with these symmetries contain contributions from both the particle momentum and the spin tensor. We denote the conserved energy and total angular momentum by $\tilde{E}$ and $\tilde{L}$, and introduce the corresponding quantities per unit mass as $E=\tilde E/m$ and $L=\tilde L/m$. For the metric considered here, they take the form
\begin{equation}
  E=-p_t-\frac{1}{2}f'(r)S^{tr},
  \qquad
  L=p_\phi+rS^{r\phi}.
  \label{eq:conserved-energy-angular-momentum}
\end{equation}
The first terms in these expressions correspond to the momentum contributions, while the additional terms arise from the coupling of the particle spin to the spacetime symmetries. In particular, \(L\) represents the conserved total angular momentum rather than the orbital angular momentum alone.

Substituting eq.~\eqref{eq:spin-components-momentum} into eq.~\eqref{eq:conserved-energy-angular-momentum}, one obtains
\begin{equation}
  p_t
  =-\frac{
  E-\dfrac{sL f'(r)}{2r}
  }{
  1-\dfrac{s^2 f'(r)}{2r}
  },
  \qquad
  p_\phi
  =\frac{
  L-sE
  }{
  1-\dfrac{s^2 f'(r)}{2r}
  }.
  \label{eq:pt-pphi-conserved}
\end{equation}

The normalization condition in eq.~\eqref{eq:momentum-normalization-1} gives the radial covariant momentum as
\begin{equation}
  p_r
  =
  \pm
  \frac{1}{\sqrt{f(r)}}
  \sqrt{
  -1
  +\frac{p_t^2}{f(r)}
  -\frac{p_\phi^2}{r^2}
  }.
  \label{eq:radial-momentum}
\end{equation}
The two signs correspond to outgoing and ingoing radial motion. The coordinate time radial equation is written as
\begin{equation}
 \dot{r} 
  =
 \frac{p^r}{p^t}.
  \label{eq:first-order-radial-equation}
\end{equation}
Together with the expressions for $p_t$ and $p_\phi$, this equation provides the radial dynamical relation used to determine circular orbits and to evaluate their linear instability.

\subsection{Lyapunov exponent}
\label{subsec:lyapunov-exponent}

The radial effective potential provides a useful framework for characterizing the local stability of particle orbits. For an unstable circular orbit, the corresponding Lyapunov exponent is determined by the behavior of the effective potential in the vicinity of the orbit and, in particular, by its second radial derivative~\cite{Cardoso:2008bp}. Following the perturbative treatment adopted in previous studies~\cite{SJeong:2023,Ciou:2025spinningRN}, we evaluate the radial instability of the spinning particle by considering a small perturbation around the circular orbit. The radial equation expressed in terms of the coordinate time can be written as

From the equations of motion obtained in the previous subsection, the radial velocity with respect to the coordinate time can be written as
\begin{equation}
 \frac{1}{2} m \dot{r}^2 + V_{\text{eff}}(r) = 0.
  \label{eq:radial-velocity-coordinate-time}
\end{equation}
where the dot denotes the derivative with respect to $t$, and the radial effective potential is defined as $V_{\rm eff}(r)=-\frac{m}{2} \dot{r}^2$. To evaluate the Lyapunov exponent, the unstable circular orbit associated with the given particle and black hole parameters must first be identified. Since $V_{\rm eff}(r)$ depends explicitly on the conserved energy $E$, the orbital radius $r_0$ cannot be determined independently of $E$. Instead, the two quantities are obtained simultaneously from the circular orbit conditions
\begin{equation}
V_{\text{eff}}(r; E, L, s) = 0, \qquad \frac{\partial V_{\text{eff}}(r; E, L, s)}{\partial r} = 0.
\label{eq:VEFF}
\end{equation}
These conditions specify the circular orbit configuration for fixed $L$, $s$, and black hole parameters. The radial stability of each solution is then determined from the local behavior of the effective potential around $r_0$. An unstable circular orbit corresponds to a local maximum of $V_{\rm eff}(r; E, L, s)$, characterized by
\begin{equation}
\left. \frac{\partial^2 V_{\text{eff}}(r; E, L, s)}{\partial r^2} \right|_{r=r_0} < 0.
\label{eq:unstable-orbit}
\end{equation}
In the subsequent analysis, only solutions located outside the event horizon, $r_0>r_h$, and consistent with the physical constraints of the spinning particle dynamics are retained. For an unstable circular orbit located at $r=r_0$, we introduce a small radial perturbation $r(t) = r_0 + \epsilon(t)$. Expanding the effective potential around \(r_0\) and retaining the leading-order perturbation gives
\begin{equation}
\ddot{\epsilon} - \lambda^2 \epsilon = 0,
\end{equation}
where the Lyapunov exponent is given by
\begin{equation}
  \lambda^2 = -\frac{\mathcal{V}''_{\text{eff}}(r_0)}{m} = \frac{1}{2} \frac{d^2}{dr^2} \left( \frac{p^r}{p^t} \right)^2 \Bigg|_{r=r_0} .
  \label{eq:unstable-circular-orbit-condition}
\end{equation}
A positive Lyapunov exponent indicates exponential sensitivity of the radial motion to small perturbations near the unstable circular orbit.

\section{Violation of the chaos bound}
\label{sec:chaos-bound-violation}

In this section, we numerically analyze the relation between the orbital Lyapunov exponent and the black-hole surface gravity. In the following calculations, we set $M=1$ and fix the primary hair charge at $l_0=1$, while the deformation parameter $\alpha$ is varied. This parameter choice follows previous studies of the same hSBH geometry, which showed that, for fixed $M=l_0=1$, changing $\alpha$ modifies the spacetime geometry and gives rise to different photon sphere configurations~\cite{Wang:2025doublephoton}. In particular, within an appropriate range of $\alpha$, the photon effective potential develops two local maxima outside the event horizon, corresponding to two unstable photon spheres~\cite{Gan:2021xdl,Gan:2021pwu,Wang:2025doublephoton}. Motivated by this structure, we investigate whether the same background also supports distinct inner and outer unstable circular orbit branches for massive spinning particles, and whether these branches exhibit different relations between the Lyapunov exponent and the surface gravity.

\begin{figure*}[t]
  \centering
  \begin{minipage}[b]{0.32\textwidth}
    \centering
    \includegraphics[width=\widthEr,height=\heightEr,keepaspectratio]{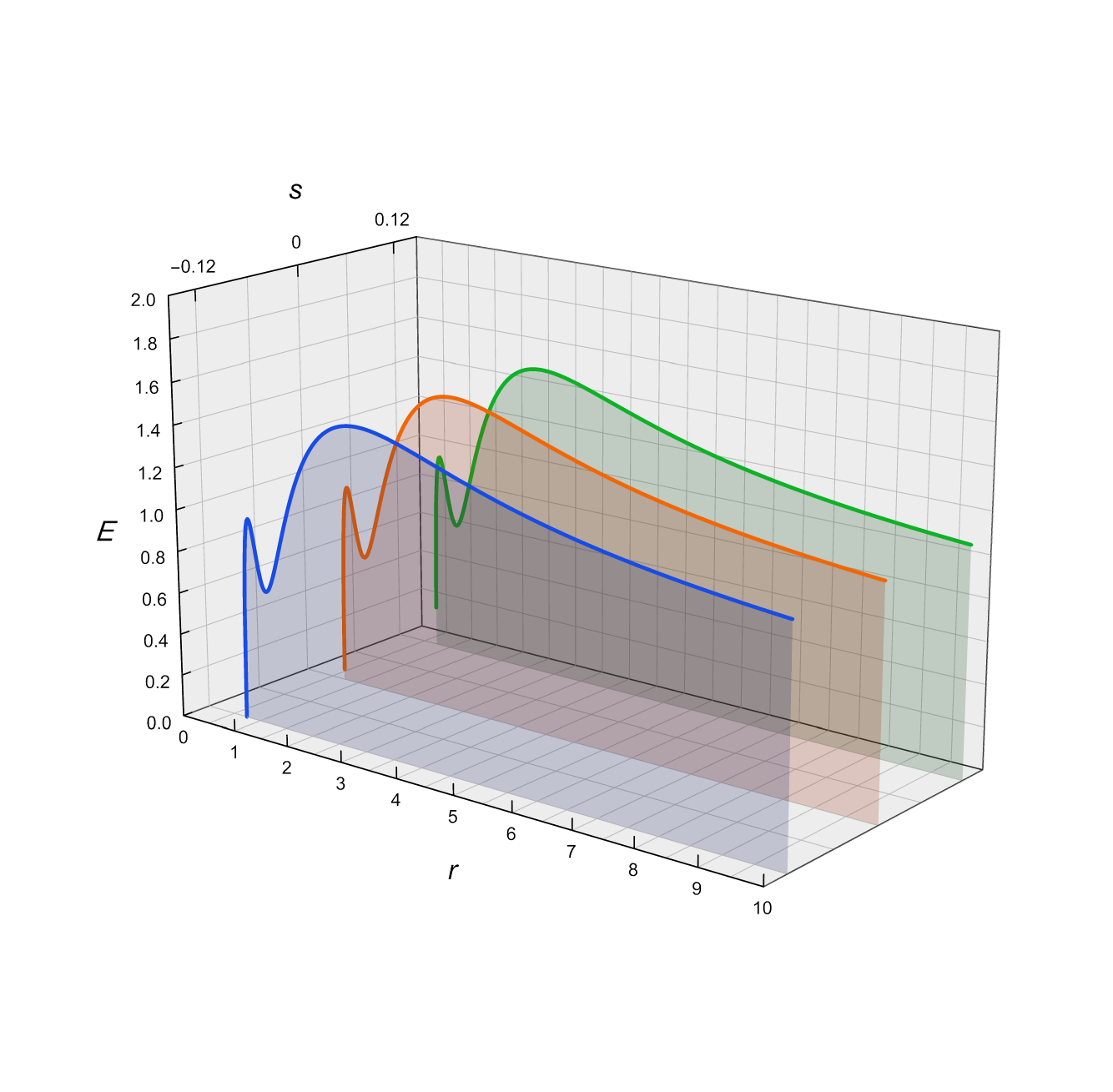}
    \panelcaption{(a) Circular orbit energy $E(r)$}
  \end{minipage}
  \hfill
  \begin{minipage}[b]{0.32\textwidth}
    \centering
    \includegraphics[width=\widthVrin,height=\heightVrin,keepaspectratio]{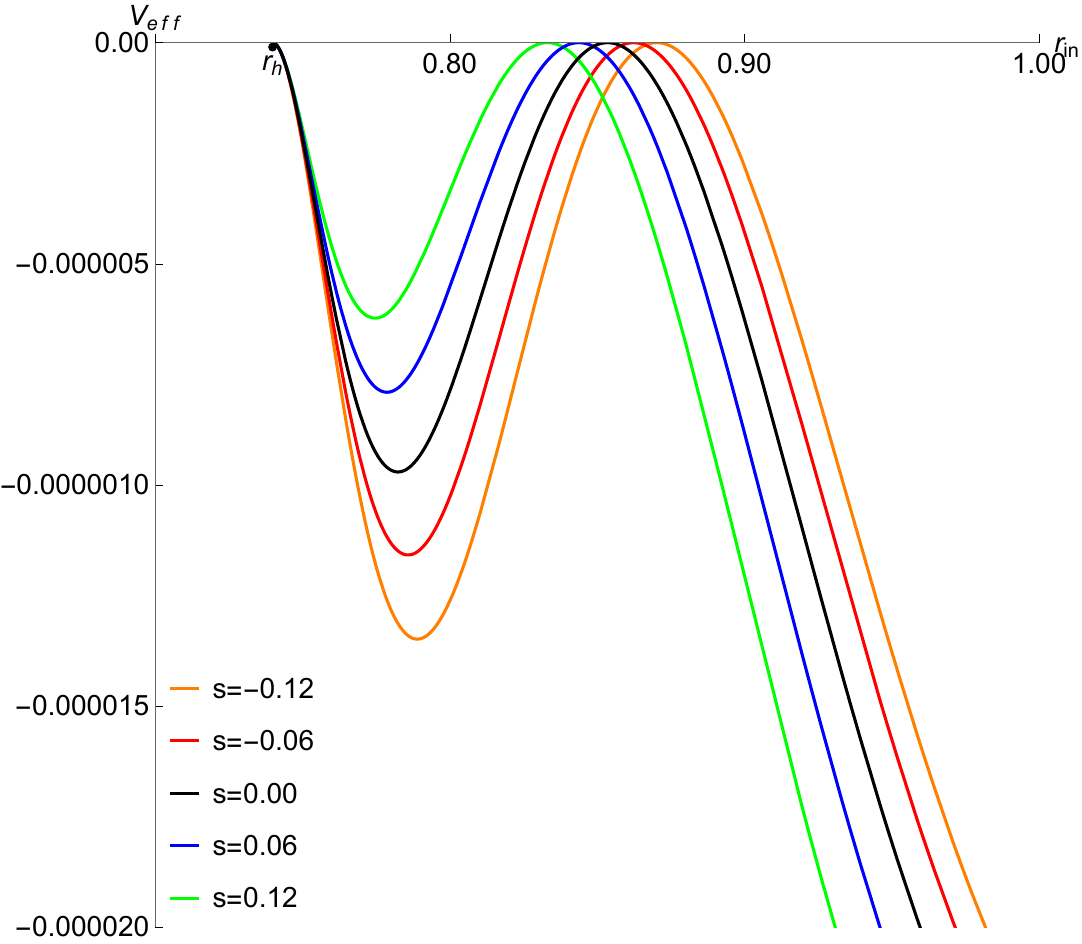}
    \panelcaption{(b) Inner orbit, $V_{\rm eff}(r)$}
  \end{minipage}
  \hfill
  \begin{minipage}[b]{0.32\textwidth}
    \centering
    \includegraphics[width=\widthVrout,height=\heightVrout,keepaspectratio]{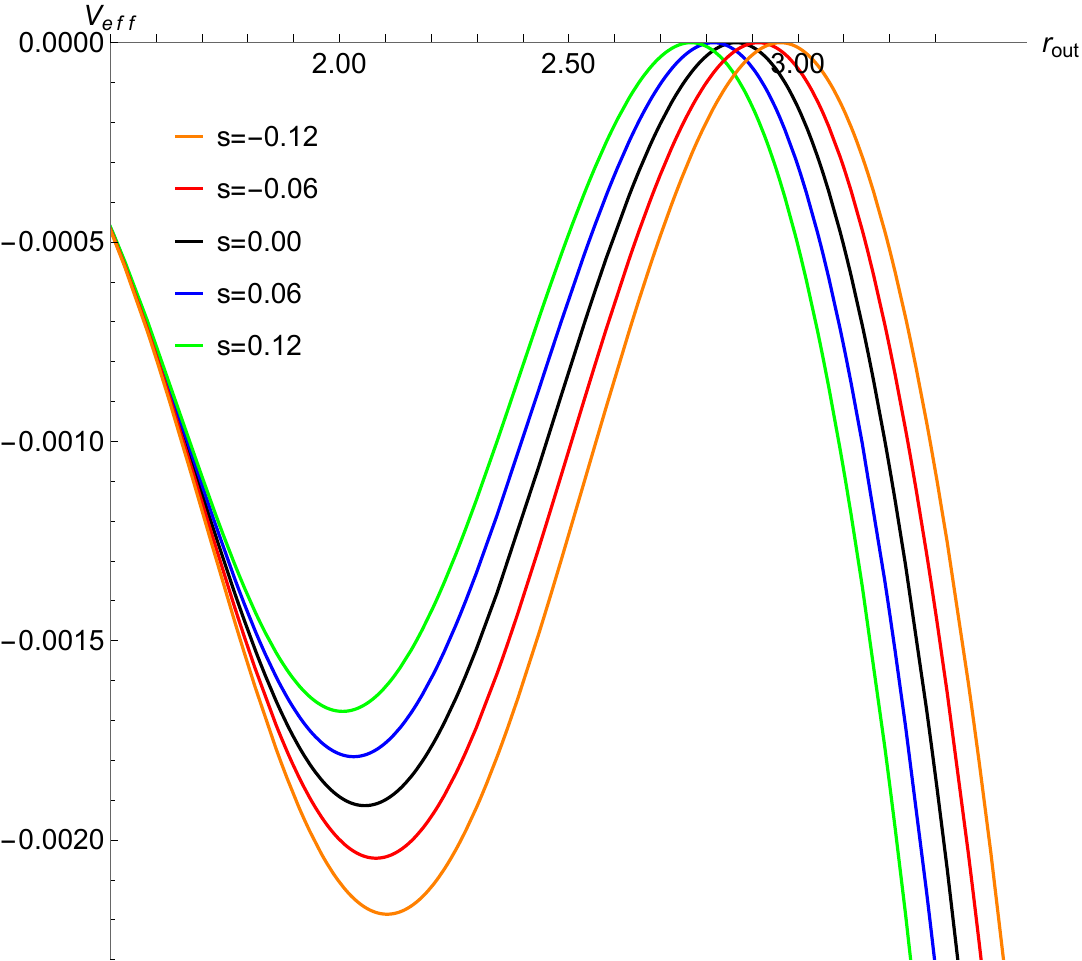}
    \panelcaption{(c) Outer orbit, $V_{\rm eff}(r)$}
  \end{minipage}
  \caption{Circular orbit structure of a spinning particle in an hSBH background. Panel (a) shows the circular orbit energy $E(r)$, while panels (b) and (c) show the local effective potential $V_{\rm eff}(r)$ in the vicinity of the inner and outer unstable circular orbits, respectively.}
  \label{fig:circular-orbit-structure}
\end{figure*}

Figure~\ref{fig:circular-orbit-structure} shows the circular orbit structure of a spinning particle and the influence of the spin parameter $s$ on the inner and outer orbit branches. Since the radial effective potential $V_{\rm eff}(r;E,L,s)$ adopted in this paper explicitly contains the particle energy $E$, the inner and outer circular orbits generally correspond to different energy values. In other words, the two circular orbits are not two extrema on the same fixed-energy effective potential curve, but instead correspond to two distinct circular orbit solutions, $(r_{\rm in},E_{\rm in})$ and $(r_{\rm out},E_{\rm out})$.
Therefore, if one fixes an energy $E$ to plot $V_{\rm eff}(r)$, it is usually not possible to clearly display the inner and outer circular orbits simultaneously in the same effective potential curve. For this reason, Panel (a) presents the circular orbit energy $E$ as a function of the radius $r$, providing a convenient representation of the overall branch structure. 
To clarify the relation between the $E(r)$ curve and the circular orbit conditions, we define the implicit function $E=E(r)$ from the equation $V_{\rm eff}(r;E(r),L,s)=0$. Differentiating the identity $V_{\rm eff}(r;E(r),L,s)=0$ with respect to $r$ gives
\begin{equation}
\frac{\partial V_{\rm eff}}{\partial r}
+\frac{\partial V_{\rm eff}}{\partial E}E'(r)=0 .
\label{eq}
\end{equation}
At a circular orbit, Eq.~\eqref{eq} further requires $\partial_rV_{\rm eff}=0$. Provided that $\partial_EV_{\rm eff}\neq0$ along the branch considered, Eq.~\eqref{eq} then gives $E'(r)=0$. Therefore, along the solution curve defined by $V_{\rm eff}=0$, the stationary points of the $E(r)$ curve coincide with the circular orbit radii. It should be stressed, however, that the extrema of $E(r)$ identify the circular orbit locations but do not by themselves determine their stability; the latter is determined by the second radial derivative of the effective potential.

It can be seen from Figure~\ref{fig:circular-orbit-structure}(a) that, for the selected parameters, the $E(r)$ curve has two local maxima. This result indicates that, in the background where a double photon sphere structure exists, the circular orbit structure of a spinning particle also exhibits an obvious double branch feature. Figures~\ref{fig:circular-orbit-structure}(b) and \ref{fig:circular-orbit-structure}(c) respectively show the local effective potentials near the inner orbit and the outer orbit. It can be seen that the effective potentials near both orbit branches behave as local maxima, so both the inner and outer orbits correspond to unstable circular orbits. This shows that, in the double photon sphere background, a spinning particle does not have only a single unstable circular orbit, but can simultaneously have inner and outer unstable circular orbit branches. This double orbit structure is the basis for the subsequent separate discussion of the exponents $\lambda_{\rm in}$ and $\lambda_{\rm out}$ of the inner and outer orbit branches. 
In addition, changes in the spin parameter shift the positions of the potential peaks near the inner and outer orbits. This indicates that the spin-curvature coupling further modulates the positions of circular orbits. However, as can be seen from the figure, the role of spin is mainly to produce corrections on the basis of the existing double orbit branches, rather than to change the double orbit structure itself. Therefore, the core result of Figure~\ref{fig:circular-orbit-structure} is that, when the background spacetime has a double photon sphere structure, spinning particles can also form inner and outer unstable circular orbit branches; the spin parameter further adjusts the specific positions of these two branches.

This double orbit structure provides an important basis for the subsequent analysis of the chaos bound problem. Since the inner and outer orbits are located in different radial regions, their corresponding local instabilities may be different. Therefore, we evaluate the Lyapunov exponents $\lambda_{\rm in}$ and $\lambda_{\rm out}$ separately and compare them with the black hole surface gravity in the following analysis.

\begin{figure*}[t]
  \centering
  \begin{minipage}[b]{0.48\textwidth}
    \centering
    \includegraphics[width=\widthLin,height=\heightLin,keepaspectratio]{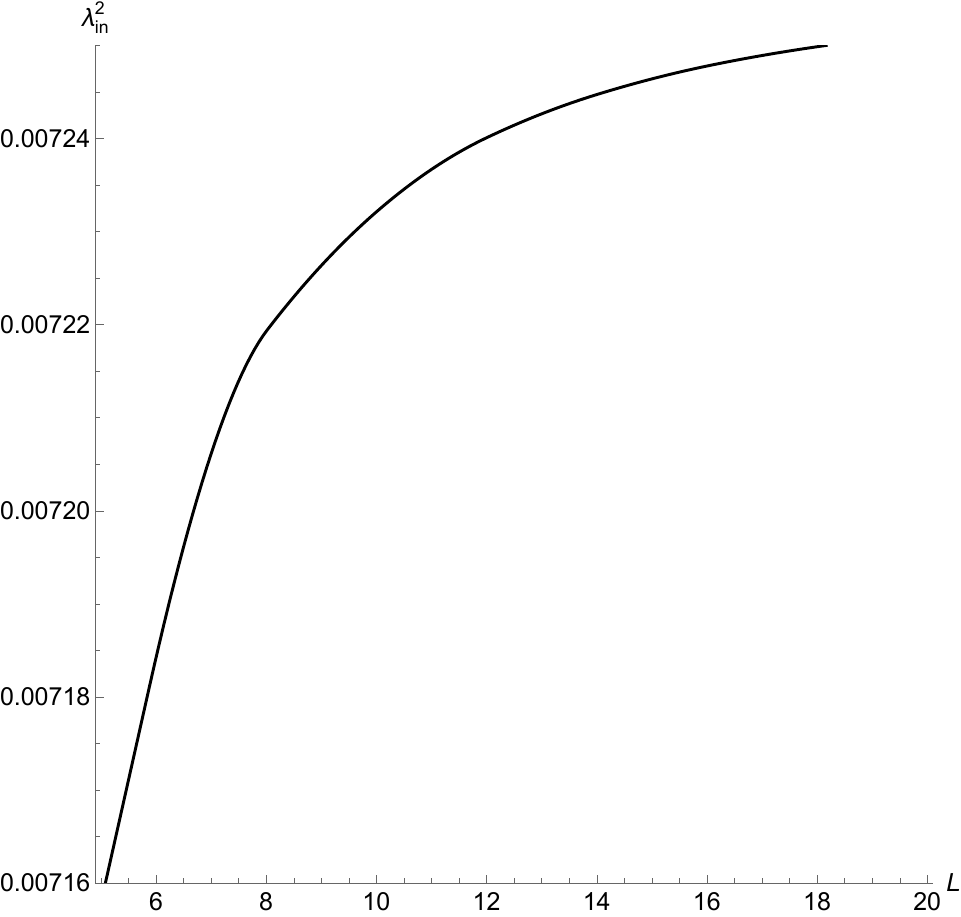}
    \panelcaption{(a) $\lambda_{\rm in}^2$ versus angular momentum}
  \end{minipage}
  \hfill
  \begin{minipage}[b]{0.48\textwidth}
    \centering
    \includegraphics[width=\widthLout,height=\heightLout,keepaspectratio]{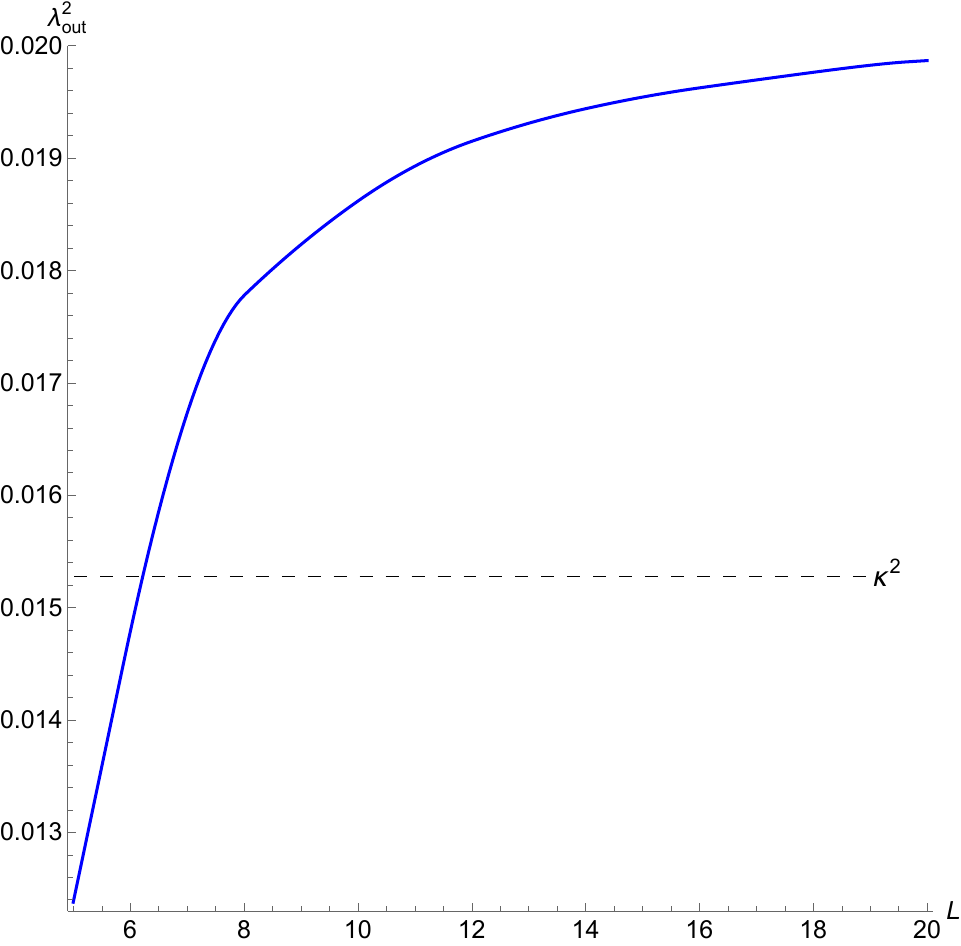}
    \panelcaption{(b) $\lambda_{\rm out}^2$ versus angular momentum}
  \end{minipage}
  \caption{Squared Lyapunov exponents on the inner and outer orbit branches as functions of the angular momentum for $s=0.16$ and $\alpha=7.48$.}
  \label{fig:lambda-vs-L}
\end{figure*}

After confirming that the particle has inner and outer unstable circular orbit branches in this case, Figure~\ref{fig:lambda-vs-L} further shows the variation of the squared Lyapunov exponent on the two branches with the particle total angular momentum. Here the spin parameter is fixed as $s=0.16$, and the deformation parameter is fixed as $\alpha=7.48$. To identify the violation of the chaos bound, the squared surface gravity, $\kappa^2$, is also shown in the figure for comparison with $\lambda^2$. When
\begin{equation}
  \lambda_i^2>\kappa^2,\qquad i={\rm in},{\rm out},
  \label{eq:chaos-bound-violation-criterion}
\end{equation}
the chaos bound is violated on the corresponding orbit branch, where $i={\rm in}$ and ${\rm out}$ label the inner and outer branches, respectively. In this case, the surface gravity of the black hole is $\kappa^2=0.015274$.
It can be seen from Figure~\ref{fig:lambda-vs-L}(a) that $\lambda_{\rm in}^2$ on the inner orbit increases as the angular momentum increases, but its overall variation is relatively small. More importantly, within the angular momentum range considered here, $\lambda_{\rm in}^2$ is always below $\kappa^2$. This shows that although the angular momentum enhances the local instability near the inner orbit, this enhancement is not sufficient to lead to a violation of the chaos bound. Therefore, the inner orbit branch always remains within the region allowed by the chaos bound in this parameter range. Different from this, Figure~\ref{fig:lambda-vs-L}(b) shows that the outer orbit branch has obviously stronger orbital instability. As the angular momentum increases, $\lambda_{\rm out}^2$ rises rapidly and crosses the $\kappa^2$ reference line within a relatively small interval of angular momentum. After the critical angular momentum is exceeded, the outer orbit satisfies $\lambda_{\rm out}^2>\kappa^2$, and thus a violation of the chaos bound occurs. Subsequently, $\lambda_{\rm out}^2$ still increases with the angular momentum, but the growth trend gradually slows down. Figure~\ref{fig:lambda-vs-L} shows that, in the double orbit structure, the chaotic properties of the inner and outer orbits are not completely the same. For the current parameters, although the inner orbit becomes more unstable as the angular momentum increases, it still remains below $\kappa^2$; while after the outer orbit exceeds the threshold value of the angular momentum, the Lyapunov exponent crosses the corresponding chaos bound threshold. Therefore, the violation of the chaos bound here does not appear simultaneously on the two orbit branches, but instead exhibits an obvious difference between orbit branches.

This result shows that the double circular orbit structure in the double photon sphere background may lead to different chaotic responses. Even if both the inner and outer orbits are unstable circular orbits, their corresponding Lyapunov exponents can have different magnitudes and growth trends. For the parameters chosen in this paper, the outer orbit responds more strongly to the angular momentum and is the first to exhibit the region where $\lambda_{\rm out}^2>\kappa^2$. This provides a basis for further examining the distribution of chaos bound violation in parameter space.

\begin{figure*}[t]
  \centering
  \begin{minipage}[b]{0.32\textwidth}
    \centering
    \includegraphics[width=\widthLsin,height=\heightLsin,keepaspectratio]{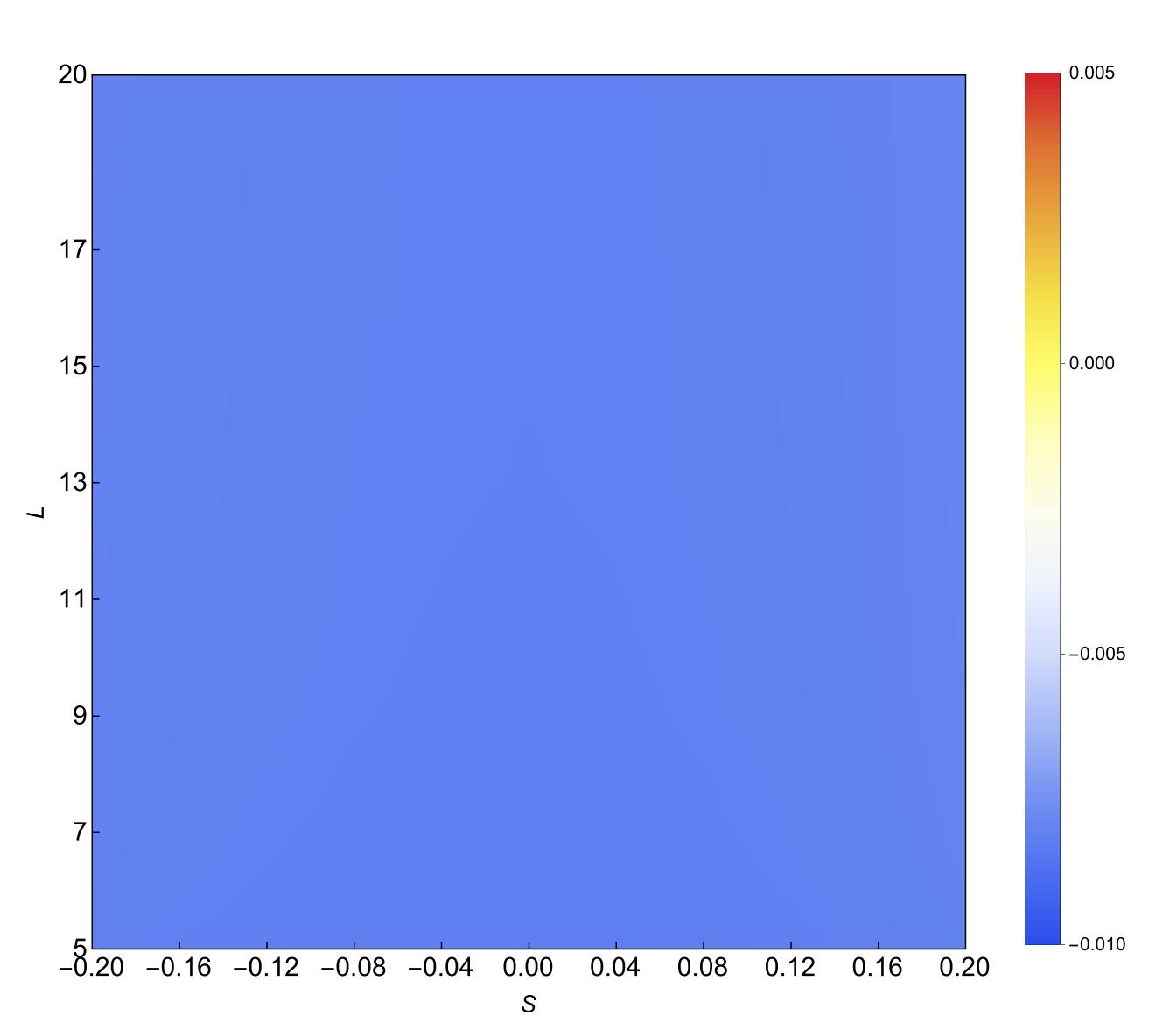}
    \panelcaption{(a) $\lambda_{\rm in}^2-\kappa^2$}
  \end{minipage}
  \hfill
  \begin{minipage}[b]{0.32\textwidth}
    \centering
    \includegraphics[width=\widthLsout,height=\heightLsout,keepaspectratio]{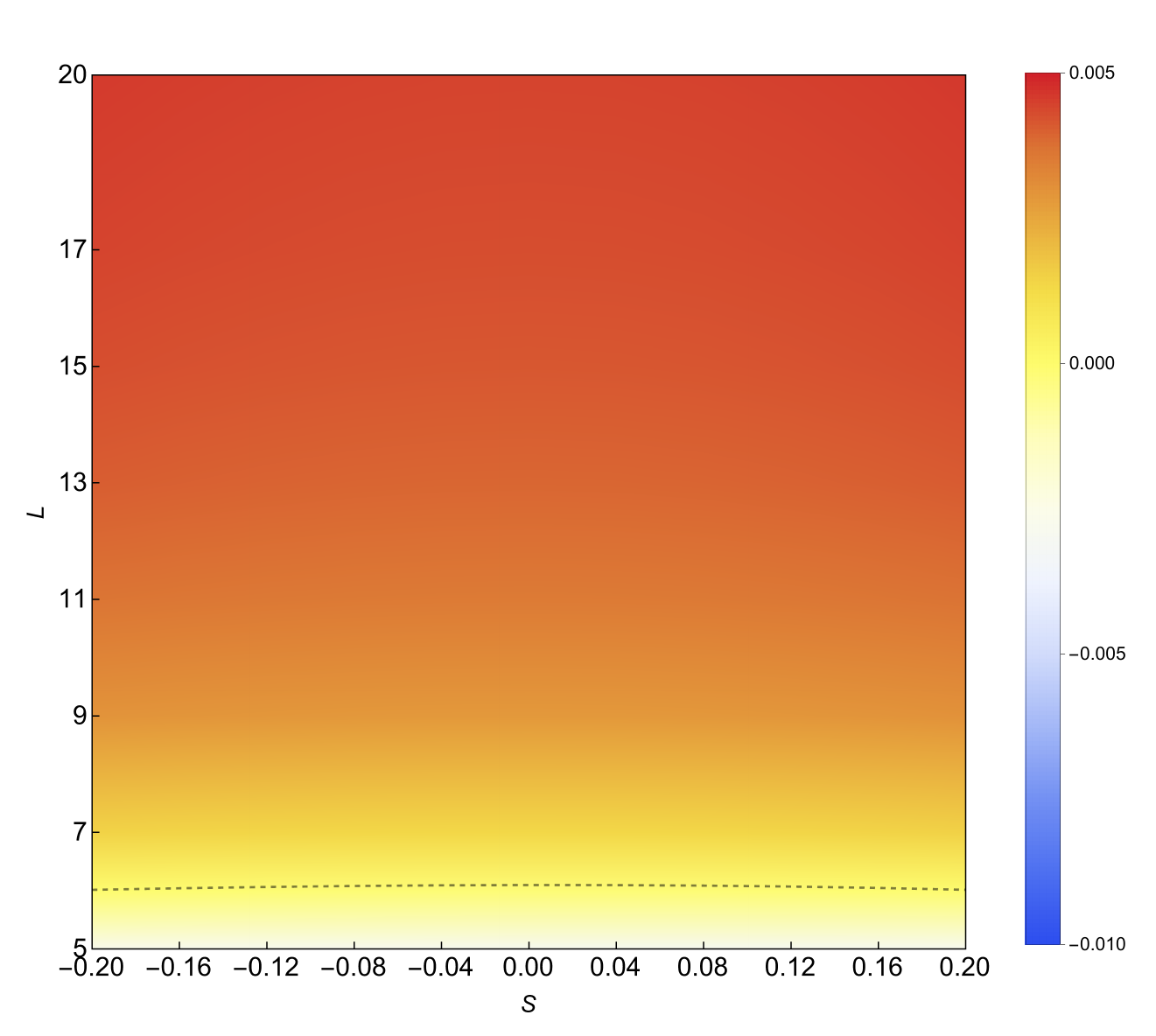}
    \panelcaption{(b) $\lambda_{\rm out}^2-\kappa^2$}
  \end{minipage}
  \hfill
  \begin{minipage}[b]{0.32\textwidth}
    \centering
    \includegraphics[width=\widthLsoutOne,height=\heightLsoutOne,keepaspectratio]{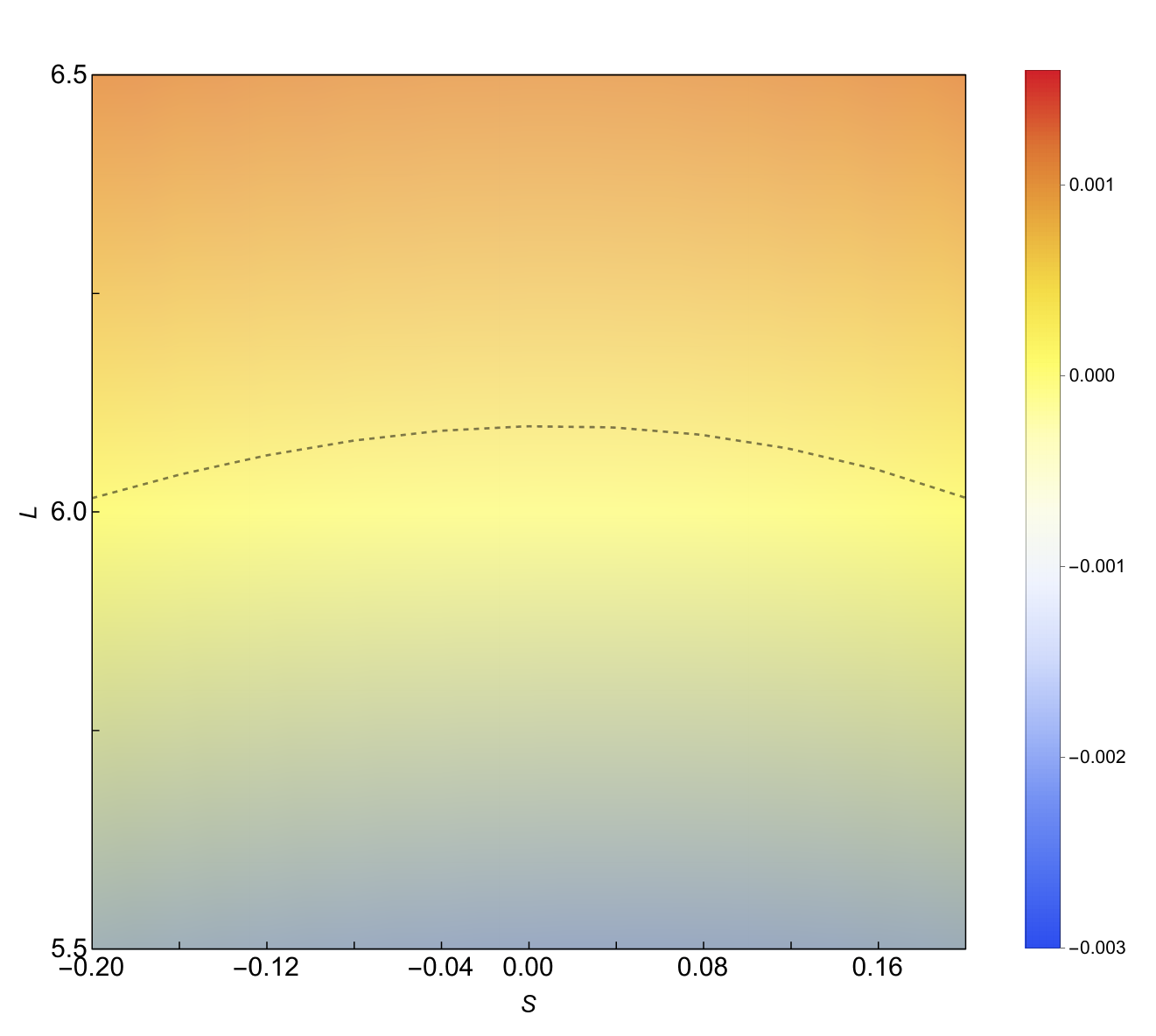}
    \panelcaption{(c) $\lambda_{\rm out}^2-\kappa^2$, local zoom}
  \end{minipage}
  \caption{Distribution of $\lambda_i^2-\kappa^2$ in the spin--angular-momentum parameter plane for the inner and outer orbit branches at fixed deformation parameter $\alpha=7.48$. The dashed curve denotes $\lambda_i^2=\kappa^2$.}
  \label{fig:lambda-s-L-plane}
\end{figure*}

Figure~\ref{fig:lambda-s-L-plane} further incorporates the spin parameter $s$ into the discussion and shows the distribution of $\lambda_i^2-\kappa^2$ in the ($s$, $L$) plane, where different colors represent the numerical magnitude of $\lambda_i^2-\kappa^2$, and the dashed line denotes the critical boundary $\lambda_i^2=\kappa^2$. Compared with Figure~\ref{fig:lambda-vs-L}, Figure~\ref{fig:lambda-s-L-plane} is no longer restricted to a fixed value of the spin, but instead shows the overall distribution of the chaos bound behavior of the inner and outer orbits under different spin parameters.
It can be seen from Figure~\ref{fig:lambda-s-L-plane}(a) that the numerical distribution corresponding to the inner orbit is overall located in the negative value region. Within the ranges of spin and angular momentum considered here, the inner orbit always satisfies $\lambda_{\rm in}^2<\kappa^2$. This shows that although both the angular momentum and the spin parameter affect the local instability of the inner orbit, within the current parameter range this influence is not sufficient to make the inner orbit cross the chaos bound. Therefore, the inner orbit still remains within the chaos bound allowed region in the two dimensional parameter space. By contrast, the outer orbit in Figure~\ref{fig:lambda-s-L-plane}(b) exhibits clearly different behavior. As the angular momentum increases, $\lambda_{\rm out}^2-\kappa^2$ gradually increases. After the angular momentum exceeds the corresponding critical value, the outer orbit enters the region $\lambda_{\rm out}^2>\kappa^2$. This indicates that, under the current parameter conditions, the violation of the chaos bound mainly appears on the outer orbit branch, rather than appearing simultaneously on both the inner and outer orbit branches. Figure~\ref{fig:lambda-s-L-plane}(c) further gives a local magnification of the critical region of the outer orbit. It can be seen that the critical line $\lambda_{\rm out}^2=\kappa^2$ is not independent of the spin parameter, but bends with $s$. This indicates that the spin-curvature coupling changes the critical condition for the outer orbit to enter the chaos bound violation region. Specifically, nonzero spin shifts the critical line toward smaller angular momentum, showing that under the same deformation parameter the spin effect can reduce the critical angular momentum required for the outer orbit to cross $\kappa^2$. Therefore, spin does not only change the numerical magnitude of the Lyapunov exponent, but also affects the boundary position of the violation region in parameter space. However, from the overall distribution, the variation of the critical line is still mainly along the angular momentum direction, while the spin parameter mainly appears as a correction to the critical boundary. This shows that, within the current parameter range, the angular momentum determines the main trend of whether the outer orbit enters the violation region, while the spin-curvature coupling further adjusts this critical process.

Therefore, Figure~\ref{fig:lambda-s-L-plane} extends the phenomenon observed at fixed spin in Figure~\ref{fig:lambda-vs-L} to the two dimensional parameter space. The inner orbit keeps $\lambda_{\rm in}^2<\kappa^2$ throughout the entire parameter region, whereas the outer orbit can exhibit $\lambda_{\rm out}^2>\kappa^2$ in the large angular momentum region. This result indicates that, in the double photon sphere background, the inner and outer unstable circular orbit branches have different chaos bound behaviors; at the same time, the spin curvature coupling further modulates the critical boundary of the violation region of the outer orbit.

\begin{figure*}[t]
  \centering
  \begin{minipage}[b]{0.48\textwidth}
    \centering
    \includegraphics[width=\widthSa,height=\heightSa,keepaspectratio]{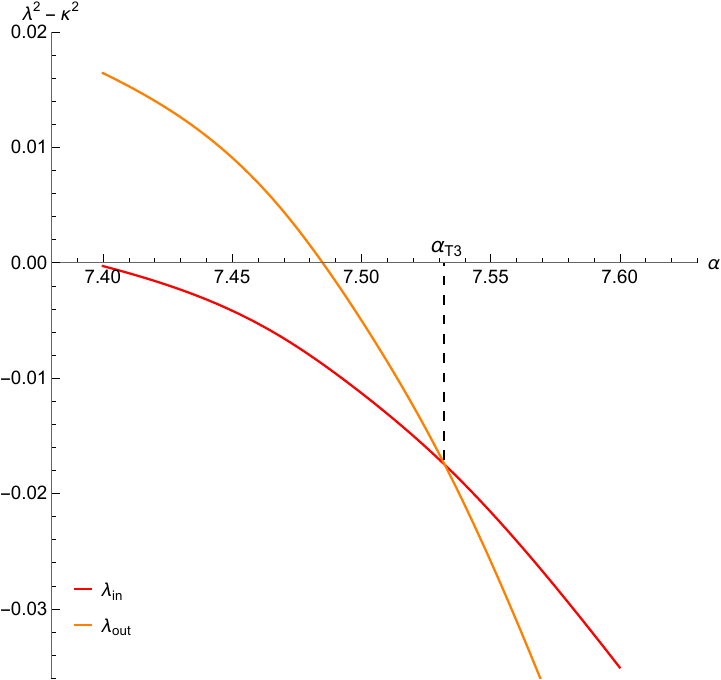}
    \panelcaption{(a) $\lambda_i^2-\kappa^2$}
  \end{minipage}
  \hfill
  \begin{minipage}[b]{0.48\textwidth}
    \centering
    \includegraphics[width=\widthSainout,height=\heightSainout,keepaspectratio]{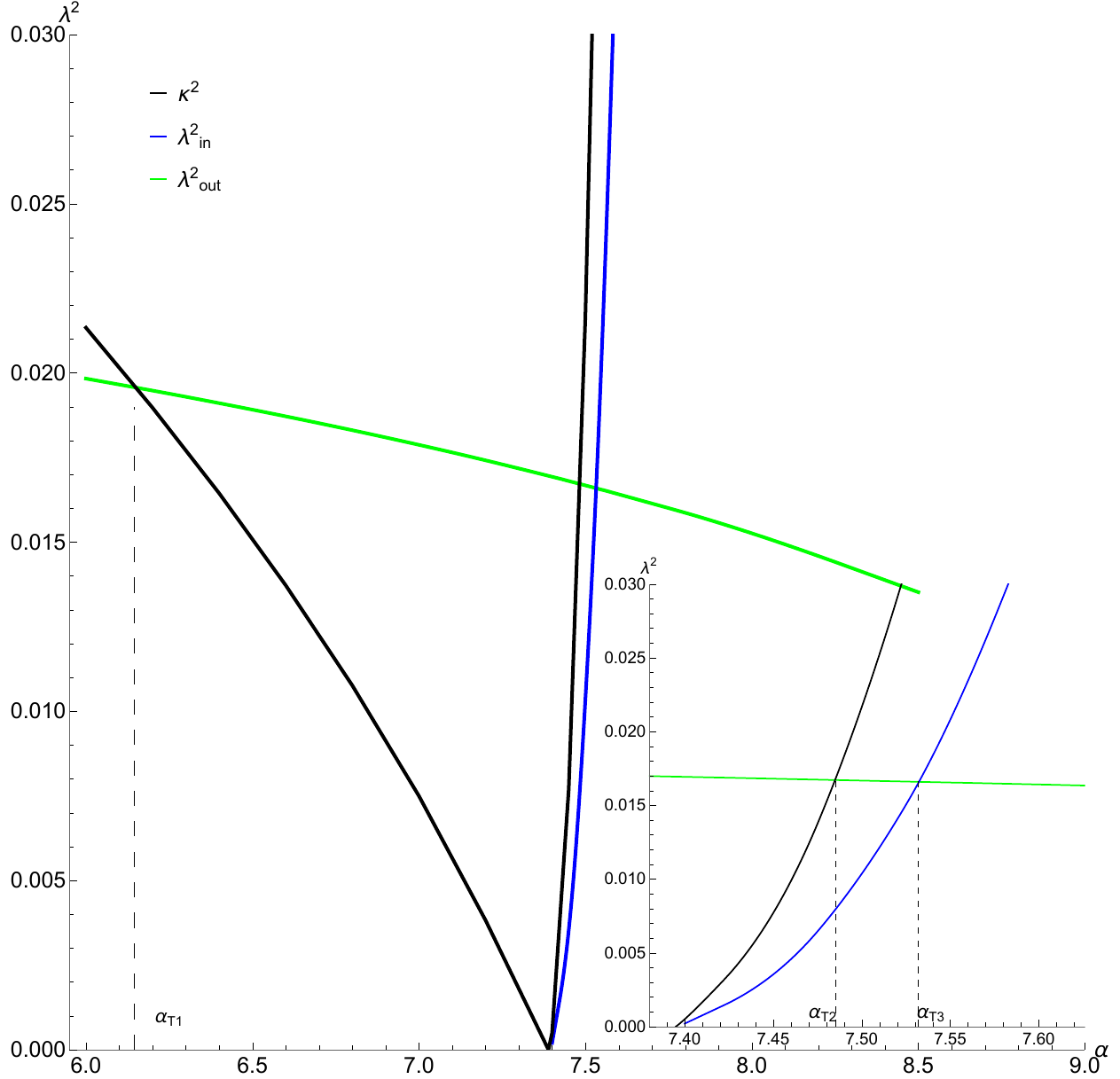}
    \panelcaption{(b) $\lambda_{\rm in}^2$, $\lambda_{\rm out}^2$ and $\kappa^2$}
  \end{minipage}
  \caption{Influence of the deformation parameter $\alpha$ on the squared Lyapunov exponents and the chaos bound.}
  \label{fig:lambda-vs-alpha}
\end{figure*}

Figure~\ref{fig:lambda-vs-alpha} illustrates how the deformation parameter $\alpha$ affects the squared Lyapunov exponents and the associated chaos bound behavior. Since this parameter belongs to the background geometry, varying it modifies both the circular orbit structure of the spinning particle and the black hole surface gravity. The behavior of $\lambda_i^2$ alone is therefore insufficient to diagnose the chaos bound; one must instead compare its evolution with that of $\kappa^2$. Accordingly, the right panel of Fig.~\ref{fig:lambda-vs-alpha} shows $\lambda_{\rm in}^2$, $\lambda_{\rm out}^2$, and $\kappa^2$, while the left panel displays $\lambda_i^2-\kappa^2$ to make the departure from the bound more transparent.
As shown in the right panel, $\lambda_{\rm out}^2$ intersects $\kappa^2$ at two critical values, $\alpha_{\tau_1}$ and $\alpha_{\tau_2}$, where $\lambda_{\rm out}^2=\kappa^2$. These crossings divide the outer branch into three regimes. For $\alpha<\alpha_{\tau_1}$, the chaos bound is satisfied; for $\alpha_{\tau_1}<\alpha<\alpha_{\tau_2}$, one has $\lambda_{\rm out}^2>\kappa^2$, and the bound is violated; for $\alpha>\alpha_{\tau_2}$, $\kappa^2$ again exceeds $\lambda_{\rm out}^2$, restoring the bound. Hence, the violation occurs only within a finite interval of the deformation parameter rather than being monotonically strengthened as the deformation increases.
The left panel focuses on the region where the two unstable branches coexist and directly shows the quantity $\lambda_i^2-\kappa^2$. For the outer branch, this quantity is initially positive but decreases rapidly with increasing deformation and crosses zero near $\alpha_{\tau_2}$. Thus, within the double branch region, increasing the deformation suppresses the chaos bound violation of the outer orbit. By contrast, the inner branch curve remains below zero throughout the displayed range, indicating that $\lambda_{\rm in}^2<\kappa^2$ and that the chaos bound is always satisfied on this branch.
The point $\alpha_{\tau_3}$ has a different physical meaning. It is defined by $\lambda_{\rm in}^2=\lambda_{\rm out}^2$ and therefore does not mark a transition between bound-satisfying and bound-violating regimes. Instead, it signals a reversal in the relative strength of the orbital instabilities: before $\alpha_{\tau_3}$, the outer branch is more unstable, whereas beyond this point the inner branch has the larger Lyapunov exponent. Since both branches satisfy $\lambda_i^2-\kappa^2<0$ in the vicinity of $\alpha_{\tau_3}$, this crossing represents a redistribution of instability between the two branches rather than the onset of a new violation region.

Figure~\ref{fig:lambda-vs-alpha} therefore reveals two distinct effects of the background deformation. First, it controls the finite interval over which the outer branch violates the chaos bound. Second, at larger deformation, the growth of the surface gravity contribution relative to the orbital Lyapunov exponent suppresses this violation. At the same time, the deformation changes the relative instability of the inner and outer branches. The resulting chaos bound behavior is thus governed by the competing evolution of the two orbital instability scales and the surface gravity, rather than by the behavior of either branch alone.

\begin{figure*}[t]
  \centering
  \begin{minipage}[b]{0.32\textwidth}
    \centering
    \includegraphics[width=\widthSain,height=\heightSain,keepaspectratio]{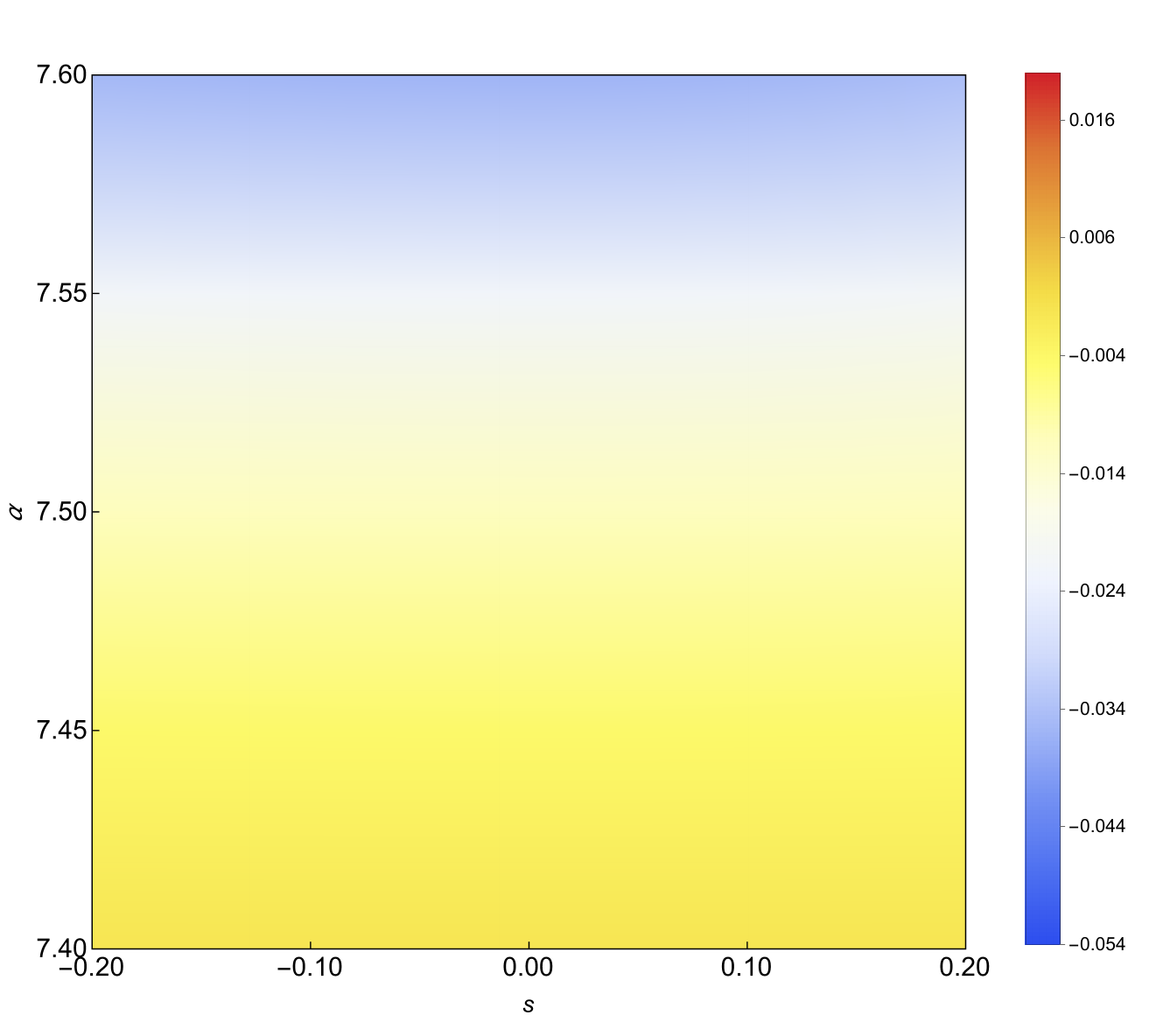}
    \panelcaption{(a) $\lambda_{\rm in}^2-\kappa^2$}
  \end{minipage}
  \hfill
  \begin{minipage}[b]{0.32\textwidth}
    \centering
    \includegraphics[width=\widthSaout,height=\heightSaout,keepaspectratio]{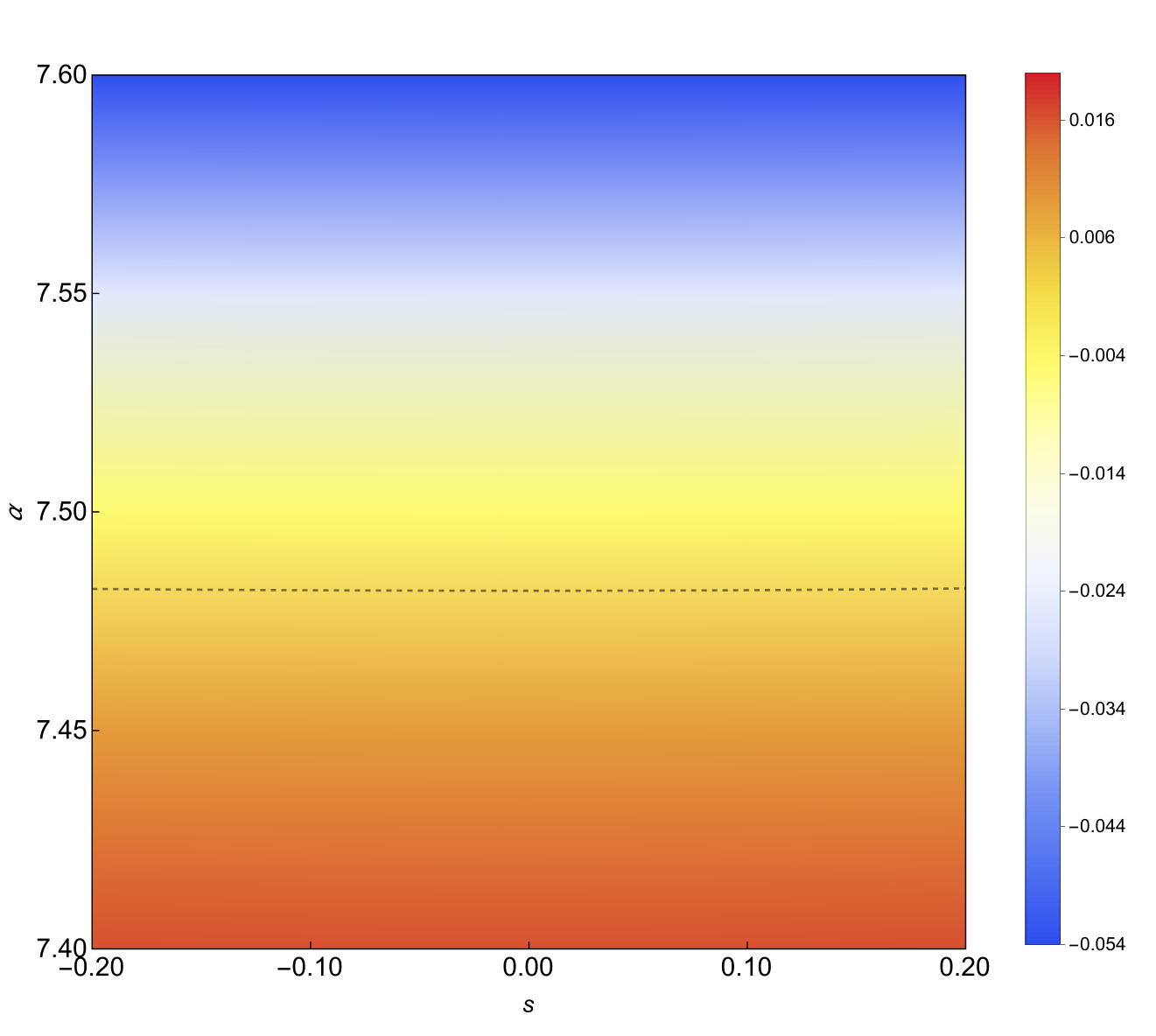}
    \panelcaption{(b) $\lambda_{\rm out}^2-\kappa^2$}
  \end{minipage}
  \hfill
  \begin{minipage}[b]{0.32\textwidth}
    \centering
    \includegraphics[width=\widthSaoutOne,height=\heightSaoutOne,keepaspectratio]{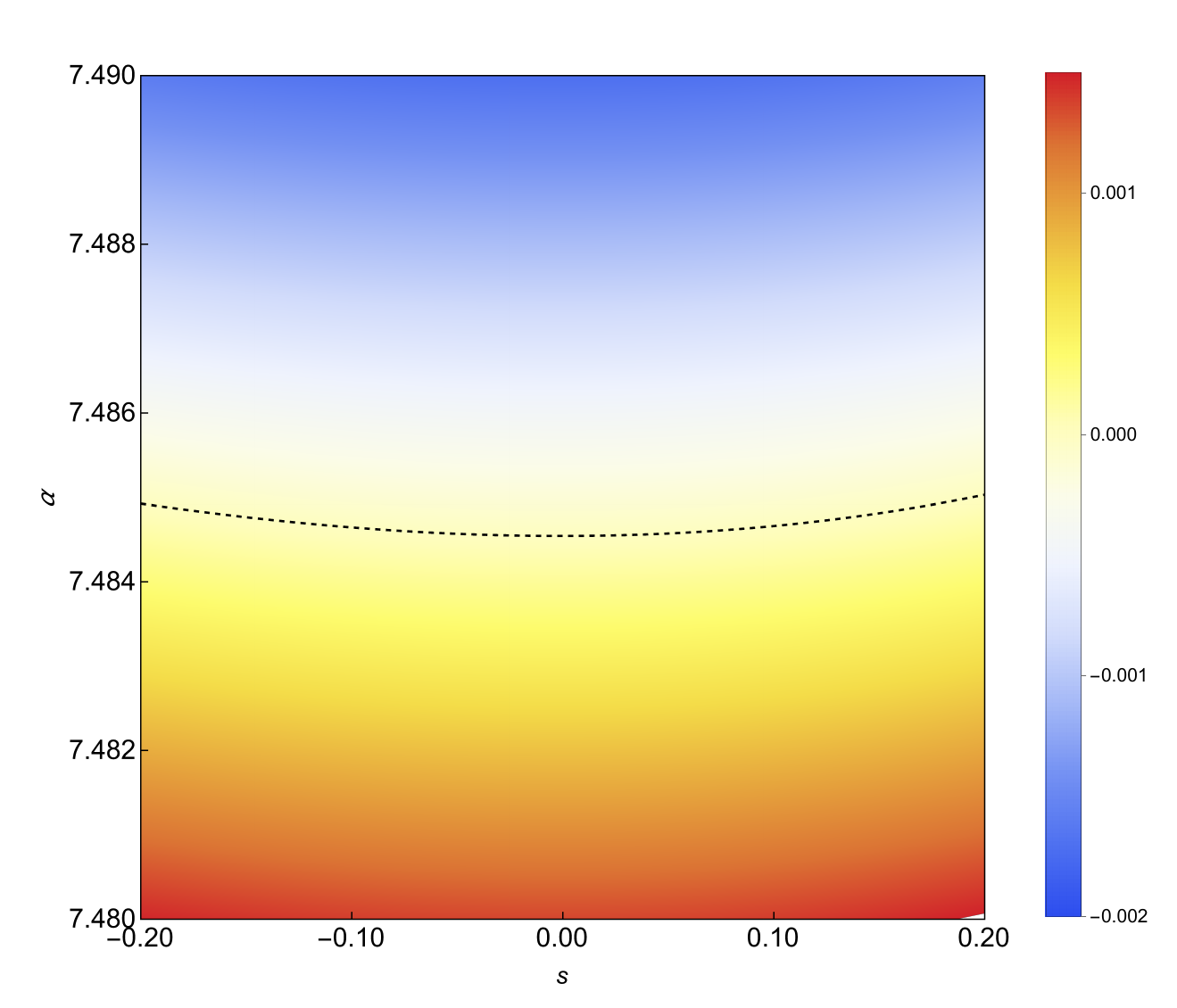}
    \panelcaption{(c) $\lambda_{\rm out}^2-\kappa^2$, local zoom}
  \end{minipage}
  \caption{Distribution of $\lambda_i^2-\kappa^2$ in the spin--deformation parameter plane for fixed angular momentum equal to 7.}
  \label{fig:lambda-s-alpha-plane}
\end{figure*}

Figure~\ref{fig:lambda-s-alpha-plane} further examines the combined influence of the deformation parameter $\alpha$ and the spin parameter $s$ on the chaos bound relation when the angular momentum is fixed at 7. Different from Figure~\ref{fig:lambda-s-L-plane}, where $\alpha$ is fixed and the ($s$, $L$) parameter plane is considered, Figure~\ref{fig:lambda-s-alpha-plane} takes the deformation parameter as the vertical axis, so that one can directly analyze how the geometric deformation modulates the chaos bound behavior of the inner and outer orbits, and whether this modulation depends on the particle spin.

It can be seen from Figure~\ref{fig:lambda-s-alpha-plane}(a) that, within the considered spin--deformation parameter range, $\lambda_{\rm in}^2-\kappa^2$ corresponding to the inner orbit always remains negative. As the deformation increases, this value decreases further overall, while its dependence on $s$ is relatively weak. This shows that, under the condition of fixed angular momentum, the spin parameter does not make the inner orbit cross the chaos bound; the inner orbit remains in the bound-satisfying region, $\lambda_{\rm in}^2<\kappa^2$. Figure~\ref{fig:lambda-s-alpha-plane}(b) shows that the behavior of the outer orbit is clearly different. For weaker deformation, the outer orbit has a region where $\lambda_{\rm out}^2-\kappa^2>0$. As the deformation becomes stronger, this value gradually decreases and crosses zero near the dashed line. This dashed line corresponds to $\lambda_{\rm out}^2=\kappa^2$, and therefore gives the critical boundary where the outer orbit moves from the chaos bound violation region back into the bound-satisfying region. This result is consistent with the conclusion from the preceding one-dimensional plots: increasing the background deformation suppresses the violation of the chaos bound on the outer orbit. Figure~\ref{fig:lambda-s-alpha-plane}(c) further magnifies the critical region of the outer orbit. It can be seen that the critical boundary is not a strictly horizontal line, but bends slightly with $s$. This shows that the spin-curvature coupling affects the boundary position of the outer orbit violation region in the spin--deformation parameter plane. Specifically, the critical line is slightly shifted toward smaller $\alpha$ near $s=0$, while the critical value of the deformation parameter increases slightly as $|s|$ increases. In other words, nonzero spin slightly extends the violation region of the outer orbit in the $\alpha$ direction, allowing the violation of the chaos bound to persist up to slightly larger values of the background deformation.

Therefore, Figure~\ref{fig:lambda-s-alpha-plane} shows that the influence of spin on the chaos bound behavior is mainly reflected in the correction to the critical boundary of the outer orbit. For the inner orbit, spin does not lead to a crossing of the bound; for the outer orbit, spin can adjust the critical position at which the background deformation suppresses the violation. Thus, when the angular momentum is fixed at 7, the deformation remains the dominant factor controlling the disappearance of the violation on the outer branch, while the spin parameter produces a further correction to this critical boundary through spin-curvature coupling. Overall, Figure~\ref{fig:lambda-s-alpha-plane} is complementary to Figure~\ref{fig:lambda-s-L-plane}. Figure~\ref{fig:lambda-s-L-plane} shows that, when $\alpha$ is fixed, increasing the angular momentum can push the outer orbit across the chaos bound; Figure~\ref{fig:lambda-s-alpha-plane} shows that, when the angular momentum is fixed, increasing the background deformation suppresses the violation behavior of the outer orbit. Together, the two figures indicate that the violation of the chaos bound on the outer orbit is not determined by a single parameter, but is jointly controlled by the particle parameters and the deformation of the background geometry. Although the spin effect is not the dominant factor, it changes the critical boundary of the violation region, thereby reflecting the interplay between particle spin and the background geometry in the double-orbit system.

\begin{figure*}[t]
  \centering
  \begin{minipage}[b]{0.32\textwidth}
    \centering
    \includegraphics[width=\widthLainPos,height=\heightLainPos,keepaspectratio]{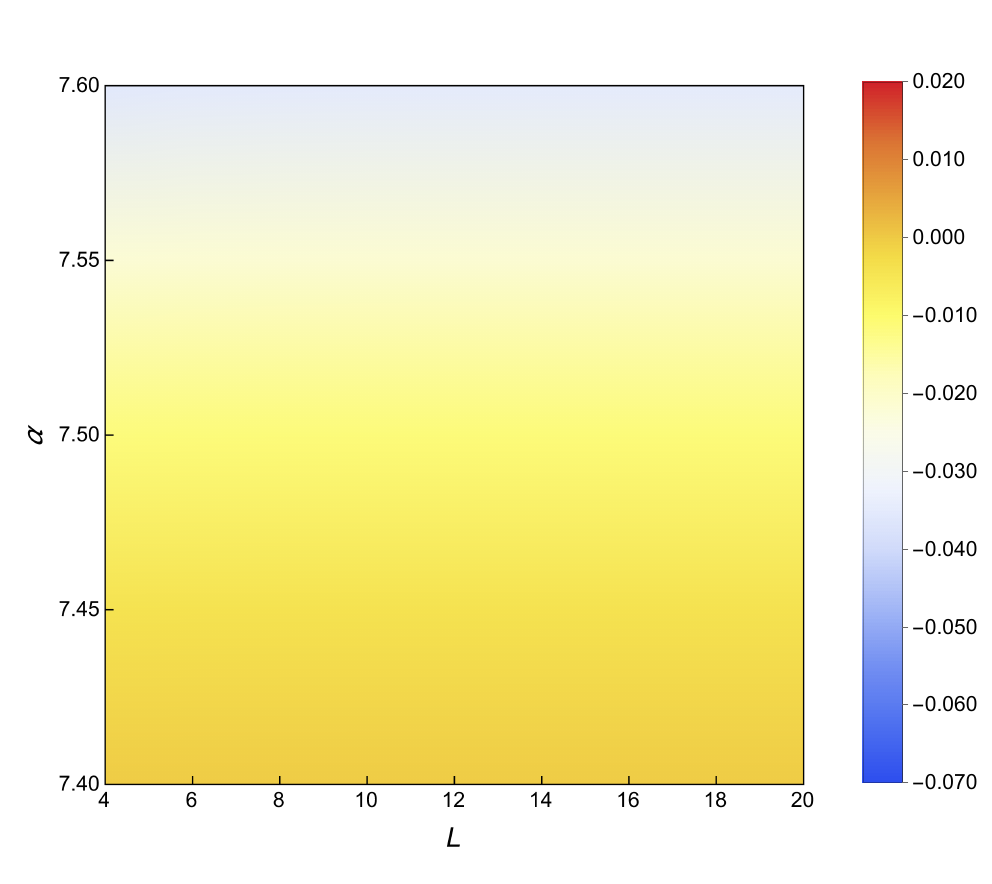}
    \panelcaption{(a) Inner orbit, $s=0.16$}
  \end{minipage}
  \hfill
  \begin{minipage}[b]{0.32\textwidth}
    \centering
    \includegraphics[width=\widthLainZero,height=\heightLainZero,keepaspectratio]{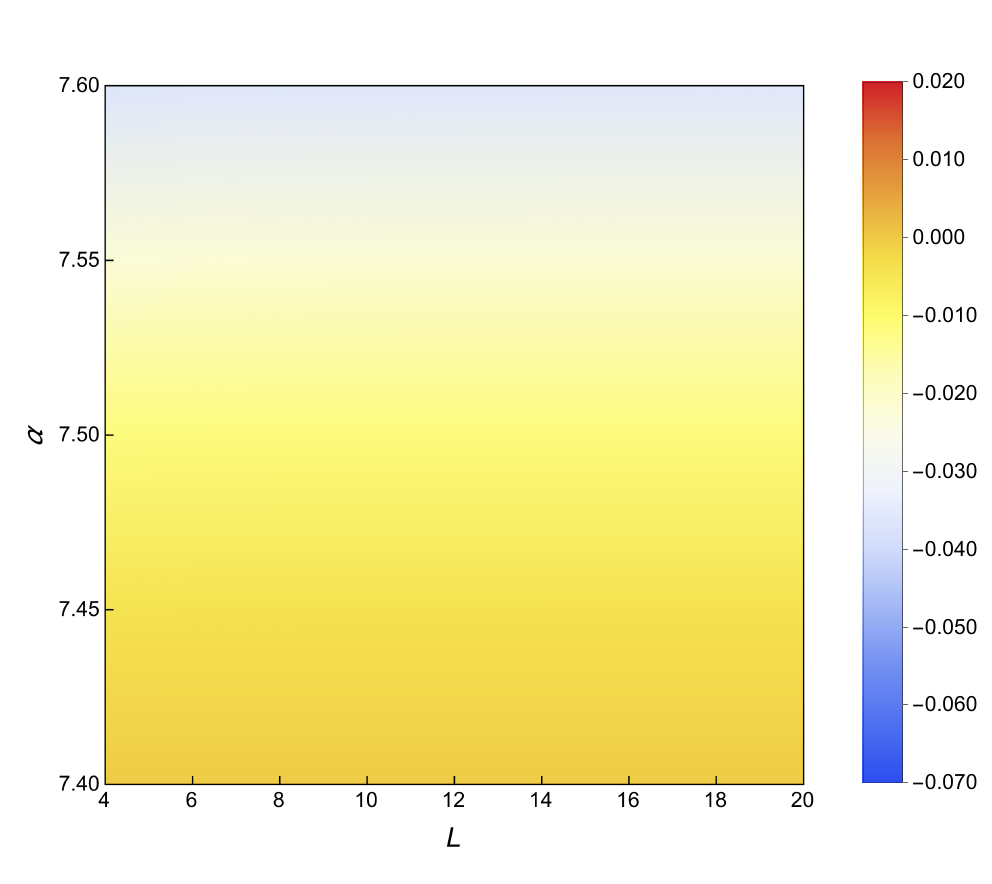}
    \panelcaption{(b) Inner orbit, $s=0$}
  \end{minipage}
  \hfill
  \begin{minipage}[b]{0.32\textwidth}
    \centering
    \includegraphics[width=\widthLainNeg,height=\heightLainNeg,keepaspectratio]{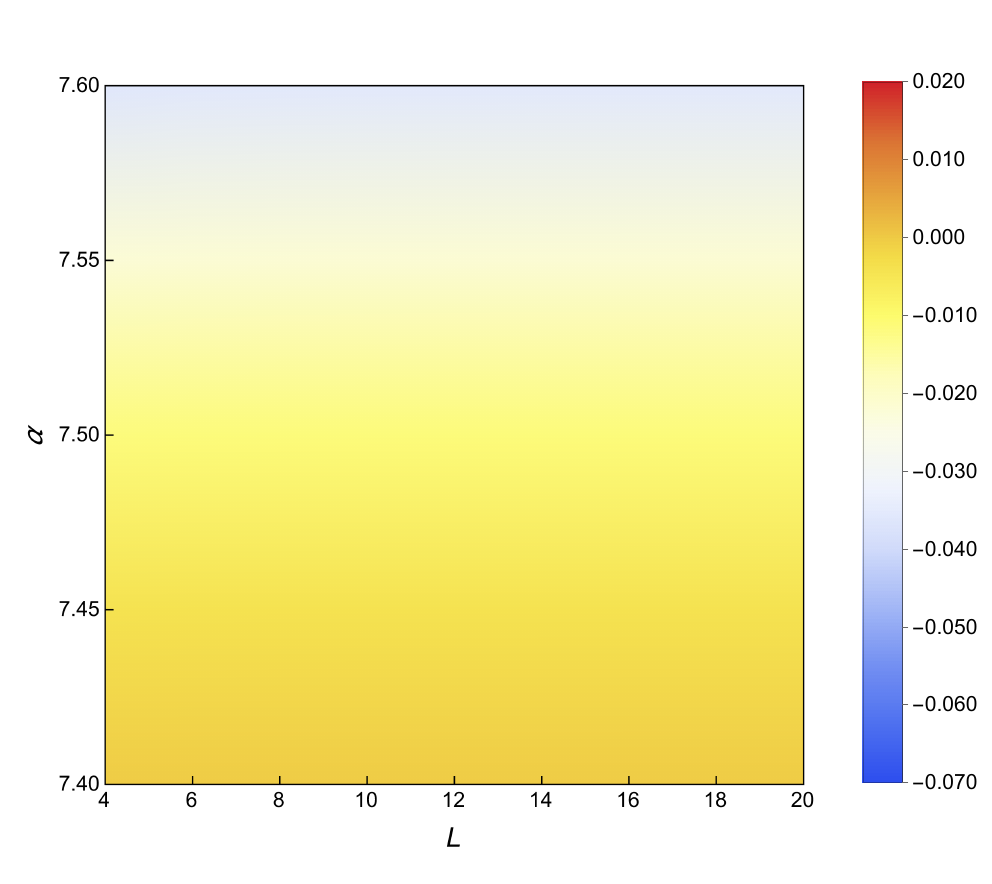}
    \panelcaption{(c) Inner orbit, $s=-0.16$}
  \end{minipage}

  \medskip

  \begin{minipage}[b]{0.32\textwidth}
    \centering
    \includegraphics[width=\widthLaoutPos,height=\heightLaoutPos,keepaspectratio]{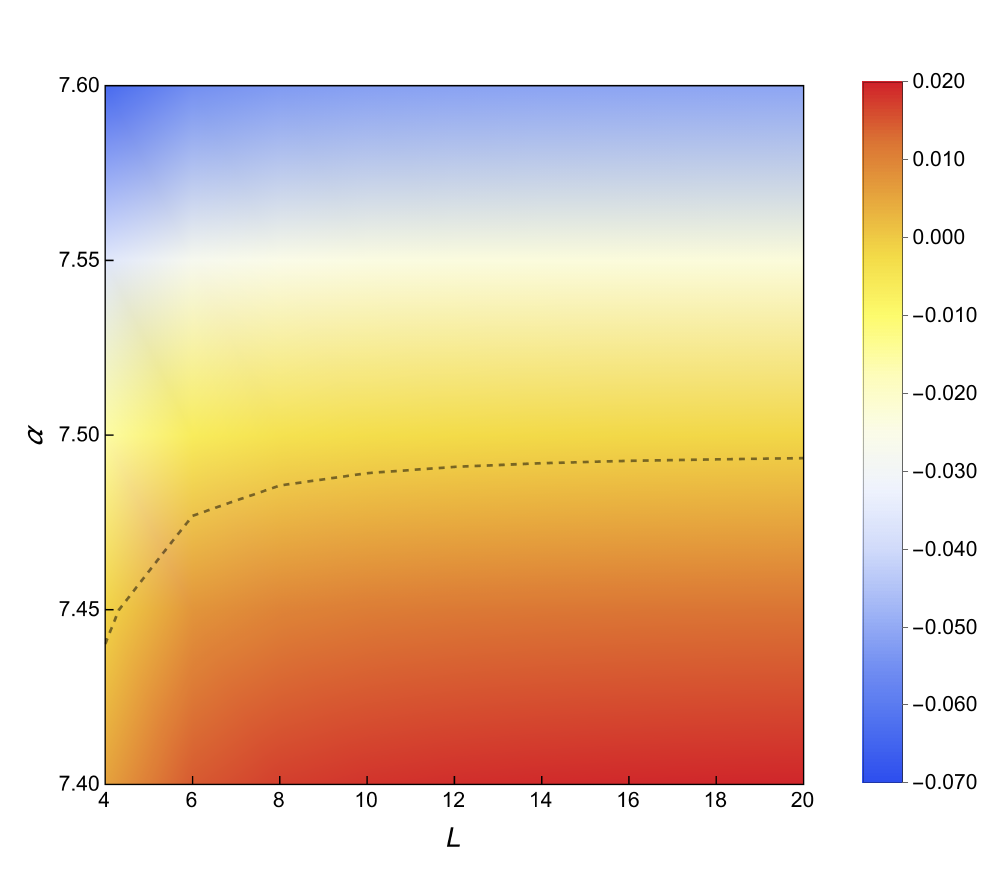}
    \panelcaption{(d) Outer orbit, $s=0.16$}
  \end{minipage}
  \hfill
  \begin{minipage}[b]{0.32\textwidth}
    \centering
    \includegraphics[width=\widthLaoutZero,height=\heightLaoutZero,keepaspectratio]{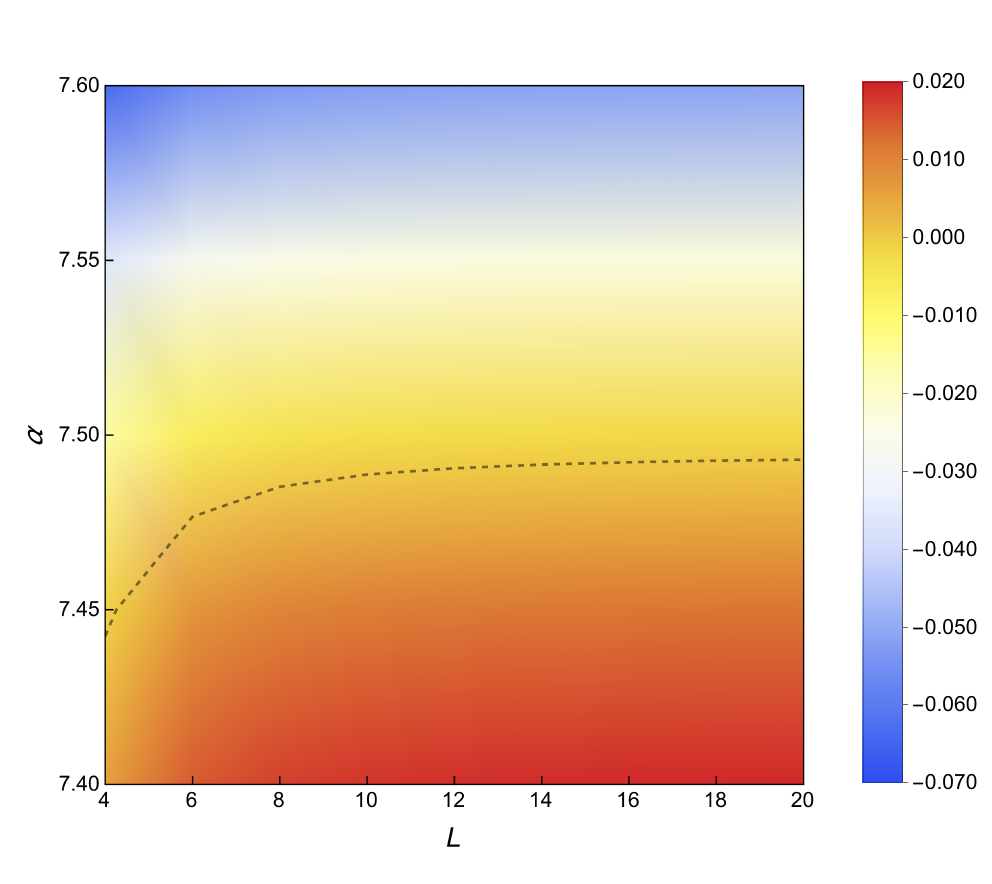}
    \panelcaption{(e) Outer orbit, $s=0$}
  \end{minipage}
  \hfill
  \begin{minipage}[b]{0.32\textwidth}
    \centering
    \includegraphics[width=\widthLaoutNeg,height=\heightLaoutNeg,keepaspectratio]{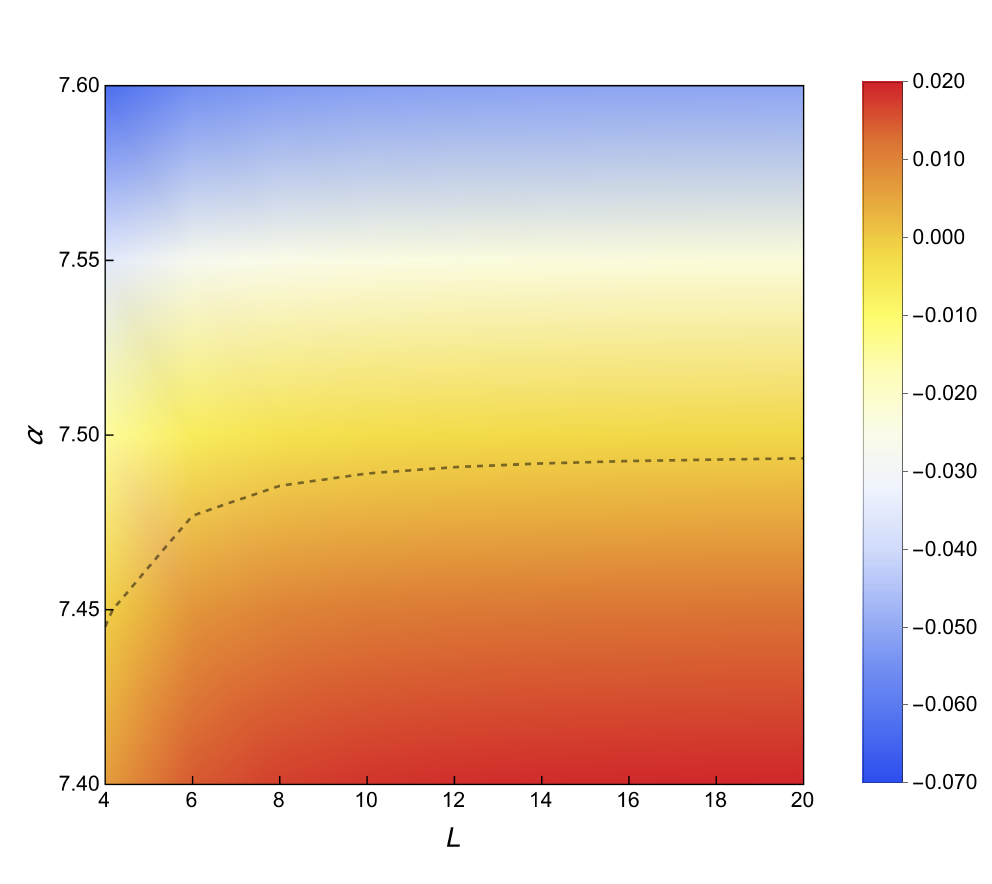}
    \panelcaption{(f) Outer orbit, $s=-0.16$}
  \end{minipage}
  \caption{Distribution of $\lambda_i^2-\kappa^2$ in the deformation--angular-momentum parameter plane for fixed spin values $s=0.16$, $s=0$ and $s=-0.16$. The upper row shows the inner orbit and the lower row shows the outer orbit.}
  \label{fig:lambda-alpha-L-plane}
\end{figure*}

Figure~\ref{fig:lambda-alpha-L-plane} further gives the distribution of $\lambda_i^2-\kappa^2$ in the deformation--angular-momentum plane for fixed spin values. The three columns from left to right correspond to $s=0.16$, $s=0$ and $s=-0.16$, respectively. The upper row gives the results for the inner orbit, and the lower row gives the results for the outer orbit. Different from the preceding one dimensional variations of angular momentum, spin or deformation parameter discussed separately, Figure~\ref{fig:lambda-alpha-L-plane} simultaneously shows the combined influence of the background deformation and the angular momentum on the chaos bound behavior.
It can be seen from the upper row that, for the three spin values, the inner orbit remains in the region $\lambda_{\rm in}^2-\kappa^2<0$ throughout the entire deformation--angular-momentum parameter region. This shows that, within the current parameter range, changing the angular momentum, the deformation parameter and the spin direction do not make the inner orbit cross the chaos bound. The numerical distribution of the inner orbit mainly varies with the deformation: as the deformation increases, $\lambda_{\rm in}^2-\kappa^2$ decreases overall, indicating that the inner branch moves farther away from the critical line $\lambda_{\rm in}^2=\kappa^2$. By contrast, the influence of the angular momentum on the inner orbit result is relatively weak and does not change the overall feature that it always remains below $\kappa^2$.
The lower row shows that the outer orbit has completely different behavior. For weaker deformation, the outer orbit can exhibit $\lambda_{\rm out}^2-\kappa^2>0$, whereas with increasing deformation, this value gradually decreases and crosses the critical boundary represented by the dashed line. The region above the dashed line corresponds to the outer orbit returning to the bound-satisfying region. Therefore, the chaos bound violation of the outer orbit does not cover the entire double orbit parameter interval, but is restricted to a region jointly determined by the deformation parameter and the angular momentum. The shape of the dashed line shows that the angular momentum has an obvious influence on the violation region of the outer orbit. As the angular momentum increases, the critical boundary as a whole moves toward larger deformation values. This means that a larger angular momentum can maintain the violation behavior of the outer orbit up to stronger background deformation. In other words, increasing the angular momentum expands the violation region of the outer orbit, whereas increasing the deformation tends to suppress it. The competition between the two determines whether the outer orbit satisfies $\lambda_{\rm out}^2>\kappa^2$.
Comparing the three columns, one finds that changing the spin value does not change the overall structure described above. For $s=-0.16$, $s=0$ and $s=0.16$, the inner orbit always remains within the chaos bound; the outer orbit always has a critical region divided by the dashed line. The influence of spin is mainly manifested as a slight modulation of the position of the critical boundary, rather than as a change in the basic shape of the violation region. This shows that, within the current parameter range, the main controlling factor for the chaos bound violation of the outer orbit is the competition between the background deformation and the angular momentum, while the spin parameter provides a further correction.

Therefore, Figure~\ref{fig:lambda-alpha-L-plane} further confirms the robustness of the preceding results in the two-dimensional parameter space: the inner orbit does not exhibit a violation of the chaos bound in the considered region, while the violation behavior of the outer orbit exists only in a finite parameter region. Larger angular momentum favors the outer orbit crossing the chaos bound, whereas stronger geometric deformation suppresses this tendency. This structure remains under different spin values, indicating that the violation region of the outer orbit is not an accidental phenomenon caused by a particular spin value, but is the result jointly controlled by the particle angular momentum and the background deformation in the double orbit system.

\section{CONCLUSIONS AND DISCUSSION}
\label{sec:conclusions}

In this paper, we studied the chaos bound problem of spinning particles in hSBHs with a double photon sphere structure. Unlike the usual situation in which the orbital dynamics is characterized by a single relevant unstable circular orbit branch, the background considered here supports two distinct unstable branches, corresponding to inner and outer circular orbits, within the parameter range of interest. Our results show that the classical orbital chaos bound in this black hole spacetime with multiple unstable circular orbit branches is intrinsically branch dependent.

Within the parameter range considered here, the inner orbit always satisfies $\lambda_{\rm in}^2<\kappa^2$, and no violation of the chaos bound appears. By contrast, the outer orbit can satisfy $\lambda_{\rm out}^2>\kappa^2$ in a certain parameter region. This result shows that chaos bound violation is not a common property of all unstable orbits in the double orbit system, but is closely related to the specific orbit branch. Increasing the angular momentum enhances the instability of the outer orbit and makes it easier for the outer orbit to approach or cross the chaos bound; the spin parameter adjusts the position of the critical boundary through spin-curvature coupling. The deformation parameter  $\alpha$ plays an important role in determining the chaos bound behavior. As a parameter of the background geometry, $\alpha$ affects not only the orbit structure, but also the black hole surface gravity. Therefore, the chaos bound behavior is determined by the relative variation between orbital instability and surface gravity. The results show that the chaos bound violation of the outer orbit exists only in a finite interval of $\alpha$; when $\alpha$ continues to increase, the enhancement of the surface gravity suppresses the violation behavior of the outer orbit and makes it return to the region $\lambda_{\rm out}^2<\kappa^2$. This shows that the black hole hair not only controls the background geometry supporting the double photon sphere structure but also regulates the appearance and disappearance of the chaos bound violation.

It should be emphasized that the analysis in this work is based on the test particle approximation and the pole-dipole approximation. Within this approximation framework, the mass and spin scale of the spinning particle are much smaller than the characteristic scale of the black hole. Therefore, the particle is regarded as a probe moving in a fixed background geometry, and its backreaction on the spacetime metric is not taken into account. The resulting region of chaos bound violation should thus be understood as a classical orbital dynamics result within the physically allowed parameter range. In addition, the Lyapunov exponent used in this work has a different physical meaning from the Lyapunov exponent in the MSS chaos bound. The Lyapunov exponent in this work is obtained from the radial perturbation equation of a classical spinning particle and describes the local growth rate of radial perturbations near an unstable circular orbit with respect to the coordinate time. In contrast, the Lyapunov exponent in the MSS chaos bound describes the exponential growth behavior of out-of-time-order correlators in finite temperature quantum many body systems. Therefore, the condition $\lambda^2>\kappa^2$ appearing in this work should be understood as a deviation of the classical orbital instability from the surface gravity scale, rather than a direct violation of the quantum chaos bound itself.

\end{document}